\documentclass[prd,12pt,preprint,tightenlines,nofootinbib,preprintnumbers]{revtex4-1}
\usepackage{color}
\usepackage{latexsym}
\usepackage{amsfonts}
\usepackage[latin1]{inputenc}
\usepackage[caption=false]{subfig}
\usepackage{graphicx}
\usepackage{mathrsfs}
\usepackage{amssymb}
\usepackage{amsmath}
\usepackage{verbatim}
\usepackage{multirow}
\usepackage{axodraw}
\usepackage{slashed}

\def\D0{\slash\!\!\!\!\!\!\!\!\!\:D0}

\newcommand{\thdm}{2HDM}
\newcommand{\model}{{\thdm}I}
\newcommand{\gfive}{\gamma^5}
\newcommand{\Pa}{P_1}
\newcommand{\Pb}{P_2}
\newcommand{\tb}{\tan\!\beta}
\newcommand{\sinb}{\sin\!\beta}
\newcommand{\cosb}{\cos\!\beta}
\newcommand{\cotb}{\cot\!\beta}
\newcommand{\rs}{\sqrt{s}}
\newcommand{\TeV}{\ \mathrm{TeV}}
\newcommand{\GeV}{\ \mathrm{GeV}}

\newcommand{\Hc}{H^\pm}
\newcommand{\MHc}{M_{\Hc}}
\newcommand{\sa}[1]{\sin\alpha_{#1}}
\newcommand{\ca}[1]{\cos\alpha_{#1}}
\newcommand{\fb}{\ \mathrm{fb}}
\newcommand{\ifb}{\fb^{-1}}
\newcommand{\re}{\mathrm{Re}}
\newcommand{\im}{\mathrm{Im}}
\newcommand{\BR}{\mathcal{B}}
\newcommand{\order}{\mathcal{O}}
\newcommand{\parbar}[1]{\overset{\text{\tiny(--)}}{#1}}
\newcommand{\nunubar}{\nu}
\newcommand{\nunubarl}[1]{\nunubar}
\newcommand{\bbbar}{\parbar{b}}

\newcommand{\TUD}{Institut f\"ur Kern- und Teilchenphysik, TU Dresden,
  01069 Dresden, Germany}

\begin{document}


\title{LHC explores what LEP hinted at:
\boldmath$CP$-violating type-I 2HDM}

\author{Wolfgang~Mader}		\affiliation{\TUD}
\author{Jae-hyeon~Park}		\affiliation{\TUD}
\author{Giovanni~Marco~Pruna}	\affiliation{\TUD}
\author{Dominik~St\"ockinger}	\affiliation{\TUD}
\author{Arno~Straessner}	\affiliation{\TUD}

\begin{abstract}
  The Large Hadron Collider is shown to have great scope for
  a light charged Higgs discovery, in the context of
  the $CP$-violating type-I two Higgs doublet model.
  This scenario with similar masses of $\Hc$ and $W$
  was suggested by the puzzling departure
  from charged current lepton universality found in the LEP data.
  With the lightest neutral Higgs mass set to $125\GeV$,
  the charged-neutral Higgs associated production mechanism
  can cause a significant excess in the $\tau\nu b\overline{b}$
  events over a vast range of $\tb$ as long as the Higgs mixing pattern
  avoids a few limiting cases.
  Thanks to the low $\Hc$ mass, the charged Higgs loop
  can play a striking role in neutral Higgs decays into $\gamma\gamma$,
  thereby compensating for a suppressed gluon-gluon fusion rate.
  The effect of scalar--pseudo-scalar mixing on loop-induced
  Higgs signals is also discussed.
\end{abstract}

\maketitle

\section{Introduction}


There are many motivations to introduce an extended Higgs sector.
Examples are, including obsolete ones,
fine-tuning in the weak-scale Higgs mass
\cite{higgs mass naturalness}
suggesting supersymmetry \cite{Fayet:1976et},
$CP$-violation 
\cite{Lee CPV,Weinberg:1976hu},
grand unification \cite{E6}, 
extended gauge symmetry \cite{Basso:2011na},
strong $CP$ problem \cite{Peccei:1977ur},
string theory \cite{Antoniadis:2000ena},
vacuum stability \cite{Gonderinger:2009jp},
light fermion mass/weak-scale hierarchy \cite{Haber:1978jt},
fermion flavour structure \cite{Babu:2004tn},
neutrino mass \cite{neutrino mass},
dark matter (see e.g.\ \cite{Deshpande:1977rw,idm}),
cosmic rays \cite{Goh:2009wg},
baryogenesis \cite{Turok:1990zg},
inflation \cite{Gong:2012ri},
novel collider phenomenology \cite{Schabinger:2005ei},
as well as
anomalies seen in precision (see e.g.\ \cite{Cheung:2001hz,Buras:2010mh})
and accelerator (see e.g.\ \cite{Park:2006gk,Ko:2011di})
experiments.
In this article, we entertain the possibility that
an intriguing outcome from the LEP experiment
is in fact pointing to an extended Higgs sector,
which may be directly inspected at the Large Hadron Collider (LHC)\@.

Specifically, we take as our framework
the two Higgs doublet model (\thdm) of type-I
with $\MHc$ in the vicinity of the $W$-boson mass.
The two main points of this paper are the following.
(1) The charged Higgs produced
in association with the lightest neutral Higgs can be
discovered as an excess in the
$pp\to \tau\nu b\overline{b}$ process at the LHC\@.
Moreover, this could happen even at the early stage with
$\sqrt{s} = 7\mbox{--}8\TeV$
in a substantial portion of the parameter space.
(2) With the lightest neutral Higgs mass assumed to be $\sim 125\GeV$,
the charged Higgs one-loop effect can enhance
$\BR(H_1\to \gamma \gamma)$ enough
to produce the excess observed at the recent LHC analyses
\cite{ATLAS:2012ad,Chatrchyan:2012tw,ATLAS:2012ae,Chatrchyan:2012tx}.
This enhancement is particularly welcome since
the $H_1$ production at hadron colliders is generically suppressed
due to Higgs mixing.

This light charged Higgs scenario was previously advocated
in~\cite{Park:2006gk}, whose contents we recap here.
The LEP Elecroweak Working Group has performed an analysis of
$W$ decay branching ratios.
In their report,
$\BR(W\rightarrow\tau\nu)$ appears to be higher than
$\BR(W\rightarrow e\nu)$ and $\BR(W\rightarrow \mu\nu)$, to the level of
2.8 standard deviations \cite{Alcaraz:2006mx}.
Should this difference be real, it would violate
the charged current lepton universality, that is predicted by the
gauge invariance of the Standard Model (SM)
and has been confirmed in many indirect tests.

As a resolution of this puzzle,
one of the authors pointed out that a significant portion of
the apparent lepton non-universality can be attributed to
production of light charged Higgs pairs which
predominantly decay to the $\tau\nu$ final states \cite{Park:2006gk}.
One of the most natural ways to realise a $\Hc$
that has mass $\sim m_W$ and is thus
light enough to be on-shell produced at LEP,
is to employ the type-I two Higgs doublet model (\model)\@.
In this model, all the $\Hc$--fermion couplings are suppressed like
$1/\tb$, the ratio of the two vacuum expectation values (VEVs).
With the choice of a high enough $\tb$, this enables $\Hc$ to escape from
the severe flavour changing neutral current (FCNC) constraints
as well as any process that probes the $\Hc$--fermion interaction
such as $Z\to b\overline{b}$, $t\rightarrow b\Hc$, and $B\rightarrow \tau\nu$.

Presuming the above LEP anomaly to be real,
the following points make this scenario more appealing.
First, invoking a light charge Higgs
(discussed also in~\cite{Dermisek light charged higgs})
is the only proposed solution
that does not spoil the other precision lepton universality tests
using lepton or meson decays \cite{Filipuzzi:2012mg}.
Second, the requirement of a light charged Higgs plus
$b \rightarrow s \gamma$ singles out type-I
out of the four types of {\thdm}s \cite{Barger:1989fj}
in which tree-level FCNC is forbidden by
a $Z_2$ symmetry \cite{Weinberg:1976hu}.
Note that these assumptions already fix
many aspects of the model that otherwise offer
more than one option, i.e.\
the Yukawa coupling structure, the charged Higgs mass,
and the viable range of $\tb$.
These combine to predict a distinct set of collider physics signatures.

In this paper, we shall consider $CP$-violation in the Higgs sector,
which can in general occur in a class of {\thdm}s.
In this way, we can explore Higgs mixing patterns
which received relatively less attention.
A practical advantage is that
one can analyse both the scalar and the pseudo-scalar Higgses
in a unified manner: switching between the two sectors
reduces to tuning the Higgs mixing angles, which determine
the composition of each neutral mass eigenstate.

Within the context of the Minimal Supersymmetric Standard Model (MSSM),
there are earlier studies of $CP$-violating Higgs phenomenology.
In particular,
the general $CP$-violating potential and mass spectrum can be found in
the paper of A.~Pilaftsis and C.~Wagner \cite{Pilaftsis:1999qt},
where the
importance of the scalar--pseudo-scalar transitions in Higgs phenomenology
was emphasised.
In the same article, the couplings of the charged Higgs
boson to $CP$-violating Higgs bosons were also discussed.

This article is organised as follows.
We begin by giving an overview of the model in the next section.
Section~\ref{sec:strategy} is a summary of our search strategy,
which leads to the main phenomenological results
presented in section~\ref{sec:res}.
Section~\ref{sec:H1} is devoted to
the diphoton signal of the lightest neutral Higgs.
We deliver our conclusion in section~\ref{sec:conclusions}.
In addition, we present:
the ghost Lagrangian for our setup of
\thdm\ in appendix~\ref{appe:b},
the relation between our parametrisation and perturbativity
in appendix~\ref{appe:perturbativity},
and differential cross-sections of our main process
in appendix~\ref{appe:diff}.

\section{Setup}
\label{sec:setup}

\subsection{Model}


Let us give a brief description of the type-I
\thdm\ that we have chosen.
The type of a \thdm\ refers to one of the multiple ways to organise
the Yukawa couplings.
There are four types in which
quarks or leptons of each charge couple to only one Higgs doublet
\cite{Barger:1989fj},
so that neutral Higgses are naturally prevented from mediating FCNC
\cite{Glashow:1976nt}.
To this end, we impose the softly broken 
$Z_2$ symmetry,
under which the two Higgs doublets $\Phi_{1,2}$ transform as
\begin{equation}
  \label{eq:Z2 flip}
  (\Phi_1, \Phi_2) \rightarrow (-\Phi_1, +\Phi_2) ,
\end{equation}
whereas the fermion fields remain unchanged.
This allows only the following Yukawa couplings,
\begin{equation}
  \label{eq:Yukawas}
  - \mathcal{L}_\mathrm{Y} =
  \overline{L}_{L} \Phi_2 Y_e e_R +
  \overline{Q}_{L} \Phi_2 Y_d d_R +
  \overline{Q}_{L} \epsilon\, \Phi_2^* Y_u u_R +
  {\rm h.c.},
\end{equation}
between $\Phi_2$ and fermions, so
all the fermions acquire mass from the VEV 
of $\Phi_2$ \cite{Haber:1978jt}.

The scalar potential including the soft $Z_2$-breaking terms reads
\begin{equation}
\label{eq:pot}
\begin{split}
V&=\frac{\lambda_1}{2}(\Phi_1^\dagger\Phi_1)^2
+\frac{\lambda_2}{2}(\Phi_2^\dagger\Phi_2)^2
+\lambda_3(\Phi_1^\dagger\Phi_1) (\Phi_2^\dagger\Phi_2)  \\
&+\lambda_4(\Phi_1^\dagger\Phi_2) (\Phi_2^\dagger\Phi_1)
+\frac{1}{2}\left[\lambda_5(\Phi_1^\dagger\Phi_2)^2+{\rm h.c.}\right] \\
&-\frac{1}{2}\left\{m_{11}^2(\Phi_1^\dagger\Phi_1)
\!+\!\left[m_{12}^2 (\Phi_1^\dagger\Phi_2)\!+\!{\rm h.c.}\right]
\!+\!m_{22}^2(\Phi_2^\dagger\Phi_2)\right\}. 
\end{split}
\end{equation}
In unitary gauge,
the Higgs doublet fields are expanded around the minimum like
\begin{equation}
\label{eq:Higgs expansion}
\Phi_1=
\left[
\begin{array}{c}
i \sinb H^+ \\
\frac{1}{\sqrt{2}} {(v_1 + \eta_1 -i \sinb \,\eta_3)}
\end{array}
\right], \quad
\Phi_2 =
\left[
\begin{array}{c}
-i \cosb H^+ \\
\frac{1}{\sqrt{2}} {(v_2 + \eta_2 + i \cosb \,\eta_3 )}
\end{array}
\right] .
\end{equation}
One can take a basis of $\Phi_{1,2}$ such that
$v_1$ and $v_2$ are both real.
Their quadrature sum is subject to the constraint,
$v_1^2 + v_2^2 = v^2$, where $v = 2^{-1/4} G_F^{-1/2}$
is the SM Higgs VEV\@.
Their ratio is parametrised by
\begin{equation}
  \label{eq:tanbeta}
  \tb = v_2 / v_1 ,
\end{equation}
in terms of the pseudo-scalar and charged Higgs mixing angle $\beta$.

We introduce $CP$-violation in the Higgs sector
by assuming $\im \lambda_5 \neq 0$ and $\im\,{m_{12}^2}\neq 0$, which
enables the neutral scalar and pseudo-scalar components
to mix together.  
The real symmetric mass-squared matrix $\mathcal{M}^2$
with respect to the basis $(\eta_1, \eta_2, \eta_3)$,
is diagonalised by the mass eigenstates
\begin{equation}
  H_i = R_{ij} \eta_j , \quad i,j = 1,2,3,
\end{equation}
like
\begin{equation}
\label{Eq:cal-M}
R{\cal M}^2R^{\rm T}=
\mathrm{diag}(M_1^2,M_2^2,M_3^2),
\end{equation}
using a $3\times 3$ orthogonal matrix $R$.
It is parametrised in terms of the mixing angles $\alpha_{1,2,3}$
in the form,
\begin{equation}
\label{eq:rotation matrix}
R=
\begin{bmatrix}
c_1\,c_2 & s_1\,c_2 & s_2 \\
- (c_1\,s_2\,s_3\!+\!s_1\,c_3) 
& c_1\,c_3\!-\!s_1\,s_2\,s_3 & c_2\,s_3 \\
- c_1\,s_2\,c_3\!+\!s_1\,s_3 
& - (c_1\,s_3\!+\!s_1\,s_2\,c_3) & c_2\,c_3
\end{bmatrix} ,
\end{equation}
where $s_i = \sa{i}$ and $c_i = \ca{i}$.
The neutral Higgs mass eigenvalues are assumed to be in the order,
\begin{equation}
\label{eq:mass ordering}
M_1^2 \le M_2^2 \le M_3^2 .
\end{equation}
This determines the following physical domain of the mixing angles
\cite{ElKaffas:2006nt}:
\begin{equation}
- \pi/2 < \alpha_{1,2,3} \le \pi/2 .
\end{equation}

The two $CP$-violating 
mass matrix elements are related by \cite{Khater:2003wq}
\begin{equation}
\label{eq:rel M13-M23}
({\cal M}^2)_{13}=\tb({\cal M}^2)_{23} ,
\end{equation}
which translates into the constraint,
\begin{equation}
\label{eq:M3}
M_3^2=\frac{M_1^2R_{13}(R_{12}\tb-R_{11})
+M_2^2R_{23}(R_{22}\tb-R_{21})}
{R_{33}(R_{31}-R_{32}\tb)}.
\end{equation}
We shall remain in the parameter volume
that is compatible with the mass ordering~(\ref{eq:mass ordering})
in conjunction with~(\ref{eq:M3}),
in order to avoid double counting of physically identical parameter sets.

Two more dimensionful quantities can be derived from
the scalar potential: the charged Higgs mass, 
\begin{equation}
  \label{eq:MHc}
  \MHc^2=\mu^2-\frac{1}{2} v^2(\lambda_4+\re\lambda_5) ,
\end{equation}
and the auxiliary parameter,
\begin{equation}
  \label{eq:mu}
  \mu^2 = \re\,m_{12}^2 / \sin 2\beta ,
\end{equation}
that sets the mass scale of charged/neutral Higgs particles except $H_1$.

We have the eight physical quantities determined from
the Higgs potential,
\begin{equation}
\label{eq:input}
M_1, M_2, \MHc, \alpha_1, \alpha_2, \alpha_3, \mu, \tb .
\end{equation}
In this work,
we opt to invert this dependency and
express the parameters appearing in~(\ref{eq:pot})
in terms of those in~(\ref{eq:input}) \cite{Khater:2003wq}.
The mass parameters 
are then given by
\begin{subequations}
  \begin{align}
    m_{11}^2&=\lambda_1 v_1^2+(\lambda_3 +\lambda_4 +\re\lambda_5 - 2 \nu)\,v_2^2 ,\\
m_{22}^2&=\lambda_2 v_2^2+(\lambda_3 +\lambda_4 +\re\lambda_5 - 2 \nu)\,v_1^2 ,\\
\im\,{m_{12}^2}&=\im{\lambda_5}v_1v_2,
\end{align}
\end{subequations}
with $\nu = \mu^2/v^2$ and
$\re\,m_{12}^2$ fixed by~(\ref{eq:mu}).
The quartic couplings are
\begin{subequations}
\label{eq:lambda}
\begin{align}
\begin{split}
    \lambda_1&=\frac{1}{c_\beta^2v^2}
[c_1^2c_2^2M_1^2
+(c_1s_2s_3+s_1c_3)^2M_2^2  \\
&+(c_1s_2c_3-s_1s_3)^2M_3^2
-s_\beta^2\mu^2],
\end{split}
\\
\begin{split}
\lambda_2&=\frac{1}{s_\beta^2v^2} 
[s_1^2c_2^2M_1^2
+(c_1c_3-s_1s_2s_3)^2M_2^2  \\
&+(c_1s_3+s_1s_2c_3)^2M_3^2-c_\beta^2\mu^2],
\end{split}
\\
\begin{split}
\lambda_3&=\frac{1}{c_\beta s_\beta v^2} 
\{c_1s_1[c_2^2M_1^2+(s_2^2s_3^2-c_3^2)M_2^2  \\
&+(s_2^2c_3^2-s_3^2)M_3^2]
+s_2c_3s_3(c_1^2-s_1^2)(M_3^2-M_2^2)\}  \\
&+\frac{1}{v^2}(2M_{H^\pm}^2-\mu^2),
\end{split}
\\
\lambda_4&=\frac{1}{v^2}
[s_2^2M_1^2+c_2^2s_3^2M_2^2+c_2^2c_3^2M_3^2
+\mu^2-2M_{H^\pm}^2],  \\
\re\lambda_5&=\frac{1}{v^2}
(-s_2^2M_1^2-c_2^2s_3^2M_2^2-c_2^2c_3^2M_3^2+\mu^2), \\
\begin{split}
\label{eq:lambda5I}
\im\lambda_5&=\frac{-1}{c_\beta s_\beta v^2}
\{c_\beta[c_1c_2s_2M_1^2-c_2s_3(c_1s_2s_3+s_1c_3)M_2^2  \\
&+c_2c_3(s_1s_3-c_1s_2c_3)M_3^2] 
+s_\beta[s_1c_2s_2M_1^2  \\
&+c_2s_3(c_1c_3\!-\!s_1s_2s_3)M_2^2
\!-\!c_2c_3(c_1s_3\!+\!s_1s_2c_3)M_3^2]\},  
\end{split}
\end{align}
\end{subequations}
where $s_\beta = \sinb$, $c_\beta = \cosb$.

In this approach, the solutions for $\lambda_{1,\ldots,5}$
may turn out to be non-perturbatively large,
unless one takes care to set the input parameters in~(\ref{eq:input})
to sensible values.
To avoid entering a nonsensical regime,
we shall check that the sizes of the Higgs quartic couplings
meet the conditions,
\begin{eqnarray}
\label{eq:perturbativity}
|\lambda_i|\le 4\pi, \quad i = 1,\ldots,5 .
\end{eqnarray}

Another theoretical constraint on the quartic couplings
that is closely related to perturbativity
arises from the tree-level unitarity in Higgs-Higgs scattering.
We also take this into account
using the formulation in~\cite{Ginzburg:2005dt}.
As we find in the numerical analysis,
this perturbative unitarity is always satisfied as long as
we require (\ref{eq:perturbativity}),
within the parameter space explored in this work.

The interactions of Higgs bosons with other particles are
described by the Lagrangian,
\begin{equation}
\begin{split}
  \label{eq:lagrangian}
  \Delta\mathcal{L} &= \sum_{i=1}^2 (D^\mu\Phi_i)^\dagger (D_\mu\Phi_i)
  + \mathcal{L}_\mathrm{Y} ,
\end{split}
\end{equation}
with $\mathcal{L}_\mathrm{Y}$ from~(\ref{eq:Yukawas}).
\begin{table}
  \centering
  \renewcommand{\arraystretch}{1.3}
  \begin{tabular}{c@{\extracolsep{3ex}}c}
    \hline
    $H_i \overline{u}_j u_j$ &
    $-i(m_{u_j}/v) (R_{i2}/\sinb - i R_{i3} \cotb \gfive)$
    \\
    $H_i \overline{d}_j d_j$ &
    $-i(m_{d_j}/v) (R_{i2}/\sinb + i R_{i3} \cotb \gfive)$
    \\
    $H^+ \overline{u}_j d_k$ &
    $\sqrt{2}\, (\cotb / v)  V_{jk}
    ( m_{u_j} P_L - m_{d_k} P_R )$
    \\
    $H_i \overline{l}_j l_j$ &
    $-i(m_{l_j}/v) (R_{i2}/\sinb + i R_{i3} \cotb \gfive)$
    \\
    $H^+ \overline{\nu}_j l_j$ &
    $- \sqrt{2}\, (\cotb / v)\,
    m_{l_j} P_R $
    \\
    \hline
    $H_i W^+_\mu W^-_\nu$ &
    $i (2 m_W^2/v) (R_{i1} \cosb + R_{i2} \sinb) g_{\mu\nu}$
    \\
    $H_i Z_\mu Z_\nu$ &
    $i (2 m_Z^2/v) (R_{i1} \cosb + R_{i2} \sinb) g_{\mu\nu}$
    \\
    $H_i H^+ W^-_\mu$ &
    $i (m_W/v) [ (R_{i1} \sinb - R_{i2} \cosb) i - R_{i3}] (p+p')_\mu$
    \\
    \hline
  \end{tabular}
  \caption{Feynman rules.
    The momentum directions are indicated in figure~\ref{fig:hHc via W}.
    The chirality projection matrices are defined by
    $P_{L/R} = (1\mp\gfive)/2$.}
  \label{tab:feynman rules}
\end{table}
One expands the covariant derivatives,
replaces $\Phi_{1,2}$ by (\ref{eq:Higgs expansion}),
and diagonalises the fermion and Higgs mass terms,
to arrive at the Feynman rules
in table~\ref{tab:feynman rules}.
They will be used to calculate and interpret the results
that we present in the following sections.

%

A remark is in order regarding the parametrisation
in the $CP$-conserving limit.
One can enforce $CP$-conservation simply by setting
$\alpha_2 = \alpha_3 = 0$.
This eliminates mixing between the $CP$-odd and the $CP$-even
degrees-of-freedom, as can be seen
in~(\ref{eq:rotation matrix}).
The mixing between $\eta_1$ and $\eta_2$ with the angle
$\alpha = \alpha_1 - \pi/2$, results in the lighter and the heavier
$CP$-even neutral eigenstates, $h$ and $H$.
The $CP$-odd neutral Higgs $A$ is identified with $\eta_3$.
In this case, the relation~(\ref{eq:rel M13-M23})
becomes trivial and therefore it no longer constrains $M^2_3$
to be determined as a function of the other parameters.
As a result, we are left with two less mixing angles
and one more free mass parameter
in the $CP$-conserving Higgs sector than in the $CP$-violating case.
Conventionally,
these seven free parameters are chosen to be:
$M_h, M_H, M_A, \MHc, \alpha, \beta, \mu$ \cite{Kanemura:2004mg}.

\subsection{Parameter space}


Having established the framework,
one should fix the range of parameters to study.
For future references,
we first collect the values which were chosen
for the reasons explained below:
\begin{equation}
  \label{eq:default}
  \MHc = 86\GeV, \quad
  M_1 = 125\GeV, \quad
  M_2 = 200\GeV, \quad
  \mu = 100\GeV, \quad
  \tb = 5.
\end{equation}
This set shall be used by default throughout the analysis
unless specified otherwise,
in combination with one of the two Higgs mixing configurations,
\begin{equation}
  \label{eq:benchmark points}
  (\sa{1},\sa{2},\sa{3}) =
  \begin{cases}
    (-0.6,0.1,0.5), & \text{benchmark point } \Pa, \\
    (\phantom{-}0.0,0.8,0.5), & \text{benchmark point } \Pb. 
  \end{cases}
\end{equation}

The charged Higgs mass is more or less predetermined by the LEP data.
The lower bound is provided by the direct search,
and the upper limit comes from the requirement that
the $\Hc$ pair production should restore lepton universality
in leptonic $W$ decays.
We take $\MHc$
which reduces the apparent discrepancy between
$\BR(W\to\tau\nu)$ and the average of $\BR(W\to e\nu)$
and $\BR(W\to \mu\nu)$
by 1~$\sigma$
\cite{Park:2006gk}.
The latest update from the OPAL collaboration reports
the direct search limit,
$\MHc > 82\GeV$, for $M_A \ge \MHc$
at 95\% confidence level (CL) \cite{Abbiendi:2008aa}.
In view of the recent evidences from ATLAS \cite{ATLAS:2012ae}
and CMS \cite{Chatrchyan:2012tx},
we set the lightest neutral Higgs mass to $125\GeV$.
The heavier Higgs mass parameter is
taken to be fairly higher than $M_1$.
Non-degenerate mass spectrum is needed for $CP$-violating Higgs mixing.
The last dimensionful parameter $\mu$ is set to
a representative value of the weak scale order.

The chosen $M_2$ is a mass at which
the SM Higgs production is strongly constrained at the LHC\@.
Not far from there lies also $M_3 \lesssim 300 \GeV$.
In this range,
the rates of the $WW$ and $ZZ$ channels
are limited to be roughly below half their SM predictions
\cite{SM Higgs exclusion}.
In our model, the upper bounds must be reinterpreted as arising from
processes mediated by $H_2$ and $H_3$
whose interactions are different from those of the SM Higgs.
For this, we require that
\begin{equation}
  \label{eq:H2 exclusion}
  {\sigma(gg \to H_{2,3} \to WW/ZZ)} < 0.5 \,
  {\sigma(gg \to h_\mathrm{SM} \to WW/ZZ)} .
\end{equation}

A potentially important constraint is provided by
the $\rho$ parameter \cite{Haber:1999zh}.
With the Higgs mass spectrum assumed in this work,
one can check that the one-loop correction from the Higgs bosons
to $\rho$ is small enough to pass the electroweak precision tests
\cite{Park:2006gk,Tmasssplit}.

For safety of a light charged Higgs, it is crucial to choose
a high enough $\tb$.
The strongest bound is given by $b \rightarrow s\gamma$,
which requires $\tb \gtrsim 4$ \cite{Park:2006gk}.
This range of $\tb$ also suppresses $\BR(t\rightarrow b H^+)$ below 4\%.
This maximal top-quark branching fraction coincides roughly with
the numerical values of the 95\% CL limits
from the light charged Higgs searches at the LHC \cite{LHC tbHc}.
However, these LHC constraints on $t\rightarrow b H^+$
are based on the assumption that $\BR(\Hc\to \tau\nu) = 1$,
and therefore become weaker in our setup where $\BR(\Hc\to \tau\nu) = 0.7$.
In principle, one could use any $\tb$ larger than the FCNC bound,
as long as it lets the charged Higgs decay inside the detector.
In practice, it is bounded from above by perturbativity
through~(\ref{eq:lambda}).
Obviously, this upper limit is a function of the other parameters.
In appendix~\ref{appe:perturbativity},
it is shown that $\mu$ can always be adjusted so that perturbativity holds.
One should not regard this as a fine-tuning but rather
an artifact of the employed parametrisation.
If one started by setting the parameters appearing in~(\ref{eq:pot})
and then derived the quantities in~(\ref{eq:input}),
then one could trivially satisfy perturbativity by keeping
$\lambda_{1,\ldots,5}$ inside the range~(\ref{eq:perturbativity}).
Even in this reverse approach, one could naturally obtain a high $\tb$,
as one can see from the example of inert doublet model
where an exact $Z_2$ symmetry leads to infinite $\tb$
\cite{Deshpande:1977rw}.

Among the three remaining Higgs mixing angles,
we scan $\alpha_1$ and $\alpha_2$ which determine
the composition of the lightest neutral Higgs.
In~(\ref{eq:rotation matrix}), one can
notice that $\sa{2}$ is the fraction of the $CP$-odd component in $H_1$.
We fix $\sa{3}$ to $0.5$ to have an intermediate level of $CP$-violation.
For illustrative purposes, we select two benchmark points,
labelled $\Pa$ and $\Pb$ in~(\ref{eq:benchmark points}).
The fraction of $A$ in $H_1$ is chosen to be small at $\Pa$
and large at $\Pb$.

\section{Light \boldmath$\Hc$ search strategy}
\label{sec:strategy}


Given the model and the input parameters, we consider
which charged Higgs production mechanism is most useful.
We first calculate the cross-sections of the following
standard channels:
\begin{subequations}
\begin{align} 
q\overline{q} &\rightarrow W \rightarrow H_1 \Hc ,
\label{eq:hHc}
\\
qq &\rightarrow jj + H_i \rightarrow jj + W \Hc ,
\\
q\overline{q} &\rightarrow W + H_i \rightarrow W + W \Hc ,
\\
q\overline{q} &\rightarrow Z + H_i \rightarrow Z + W \Hc ,
\\
gg &\rightarrow H_i \rightarrow W \Hc ,
\\
\label{eq:tHc}
gb &\rightarrow t H^- ,
\\
\label{eq:tbHc}
gg,q\overline{q} &\rightarrow t\overline{b} + H^+ .
\end{align}
\end{subequations}
The last two entries are meant to imply their conjugate processes as well.
The final state in~(\ref{eq:tbHc}) can be made through either
top pair production or gluon-gluon fusion \cite{DiazCruz:1992gg}.

\begin{figure}
  \subfloat[$\Pa$, $\rs=7\TeV$]{ 
  \label{sp_tb1_c1_07}
  \includegraphics[angle=0,width=0.48\textwidth ]{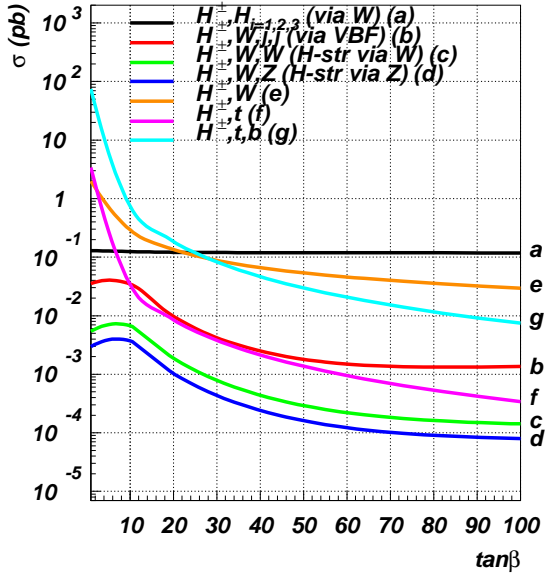}}
  \subfloat[$\Pb$, $\rs=7\TeV$]{
  \label{sp_tb1_c2_07}
  \includegraphics[angle=0,width=0.48\textwidth ]{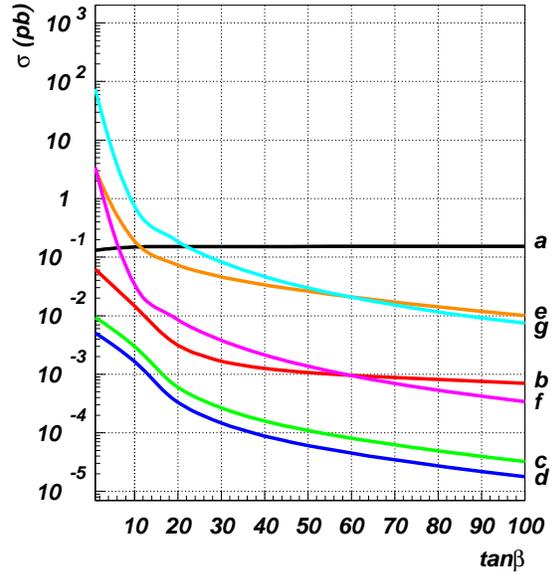}}  
\\ \vspace{-0.3cm}
  \subfloat[$\Pa$, $\rs=14\TeV$]{
  \label{sp_tb1_c1_14}
  \includegraphics[angle=0,width=0.48\textwidth ]{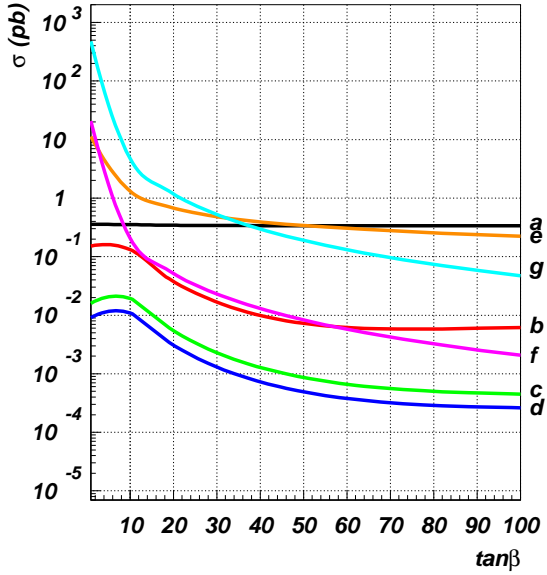}}
  \subfloat[$\Pb$, $\rs=14\TeV$]{
  \label{sp_tb1_c2_14}
  \includegraphics[angle=0,width=0.48\textwidth ]{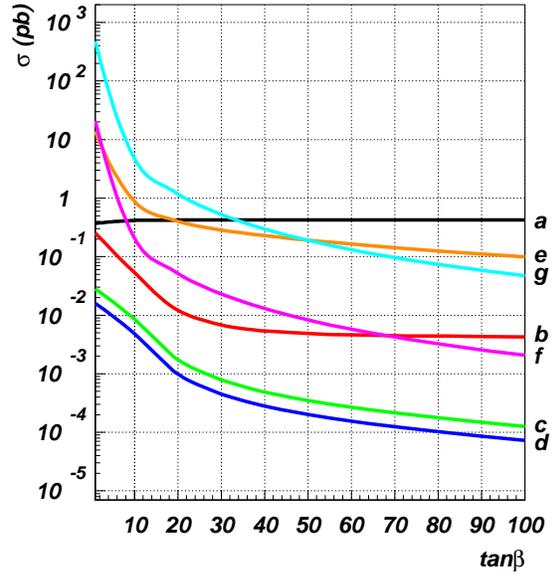}} 
\caption{Cross-sections of the single charged Higgs production mechanisms plotted against $\tb$
for the selected benchmark points and centre-of-mass energies.
The curves are independent of $\mu$.}
\label{fig:sp}
\end{figure}
In figures~\ref{fig:sp},
we plot the cross-sections against $\tb$ for the two benchmark points.
One observes the general tendency that each production rate is depleted
for high $\tb$.
As one can expect from table~\ref{tab:feynman rules},
processes~(\ref{eq:tHc}) and (\ref{eq:tbHc}) die out as $\tb$ grows,
while the others approach their individual asymptotic cross-sections
even though many of them are small.

\begin{figure}
  \centering
        \begin{picture}(180,120)(0,-60)
        \SetWidth{0.8}
        \SetColor{Black}
        \Line(25,35)(50,0)
        \Line(25,-35)(50,0)
        \Photon(50,0)(100,0){3}{4.5}
        \DashArrowLine(140,40)(100,0){4}
        \DashArrowLine(100,0)(140,-40){4}
        \Text(75,8)[b]{{{$W$}}}
        \Text(127,12)[l]{{{$p$}}}
        \Text(127,-12)[l]{{{$p'$}}}
        \Text(148,40)[l]{{{$H_{1,2,3}$}}}
        \Text(148,-40)[l]{{{$H^+$}}}
        \Text(17,35)[r]{{{$u$}}}
        \Text(17,-35)[r]{{{$\bar{d}$}}}
      \end{picture}%
  \caption{Feynman graph for the charged Higgs production in association
    with a neutral Higgs.}
  \label{fig:hHc via W}
\end{figure}
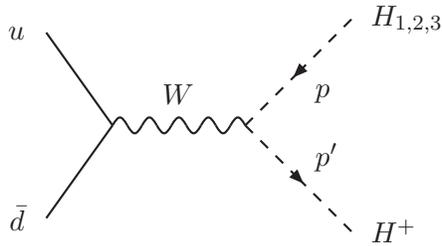
An outstanding exception is
the production associated with a neutral Higgs~(\ref{eq:hHc}),
which is not suppressed even for high $\tb$.
The diagrams for this process are shown in figure~\ref{fig:hHc via W},
which scale according to the $H_i$--$\Hc$--$W$
vertices in table~\ref{tab:feynman rules}.
In the $H_1$--$\Hc$--$W$ interaction,
the pseudo-scalar coupling is proportional to $\sa{2}$
and independent of $\beta$,
and the scalar coupling is proportional to $\ca{1}\ca{2}$
in the limit of $\tb\rightarrow\infty$.
Combining these two behaviours, one can expect that
the $H_1\Hc$ production becomes efficient unless
$|\sa{1}|$ is large and $|\sa{2}|$ is small.

The results plotted in figures~\ref{fig:sp} do not
depend on $\mu$.  Therefore, they remain valid
even if one adjusts $\mu$ for the perturbativity of $\lambda_{1,\ldots,5}$
in the high $\tb$ regime.
This means that the production mechanism~(\ref{eq:hHc}) is of the most
general interest among the displayed channels,
and we shall focus on it in what follows.
Note that the same type of diagram was previously considered in
the contexts of the MSSM
\cite{Djouadi:1999rca} and a fermiophobic Higgs scenario
\cite{Akeroyd fermiophobic higgs}.

The next step should be to select the decay products of $\Hc$ and $H_1$.
The two dominant branching ratios of $H^+$ are
\begin{equation}
  \label{eq:BRs of H+}
  \BR(H^+ \rightarrow \tau^+\nu_\tau) = 0.71, \quad
  \BR(H^+ \rightarrow c\bar{s}) = 0.27 ,
\end{equation}
as long as 
$\MHc \lesssim 135\GeV$,
beyond which the charged Higgs decays mediated by a virtual top
becomes non-negligible \cite{Ginzburg:2012hc}.
These branching fractions are essentially independent of
any of the input parameters.
This is a notable feature of the \model,
stemming from the universal scaling behaviour of
the $\Hc$--fermion couplings:
they are all proportional to $\cotb$,
as shown in table~\ref{tab:feynman rules}.
As we are mainly interested in a charged Higgs that is light enough
to have been produced at LEP energies,
we can regard (\ref{eq:BRs of H+}) as good approximations
and simply choose the decay product with the highest rate, $\tau\nu$.

\begin{figure}
 \subfloat[SM]{ 
  \label{BRH1_tb1}
   \includegraphics[angle=0,width=0.48\textwidth ]{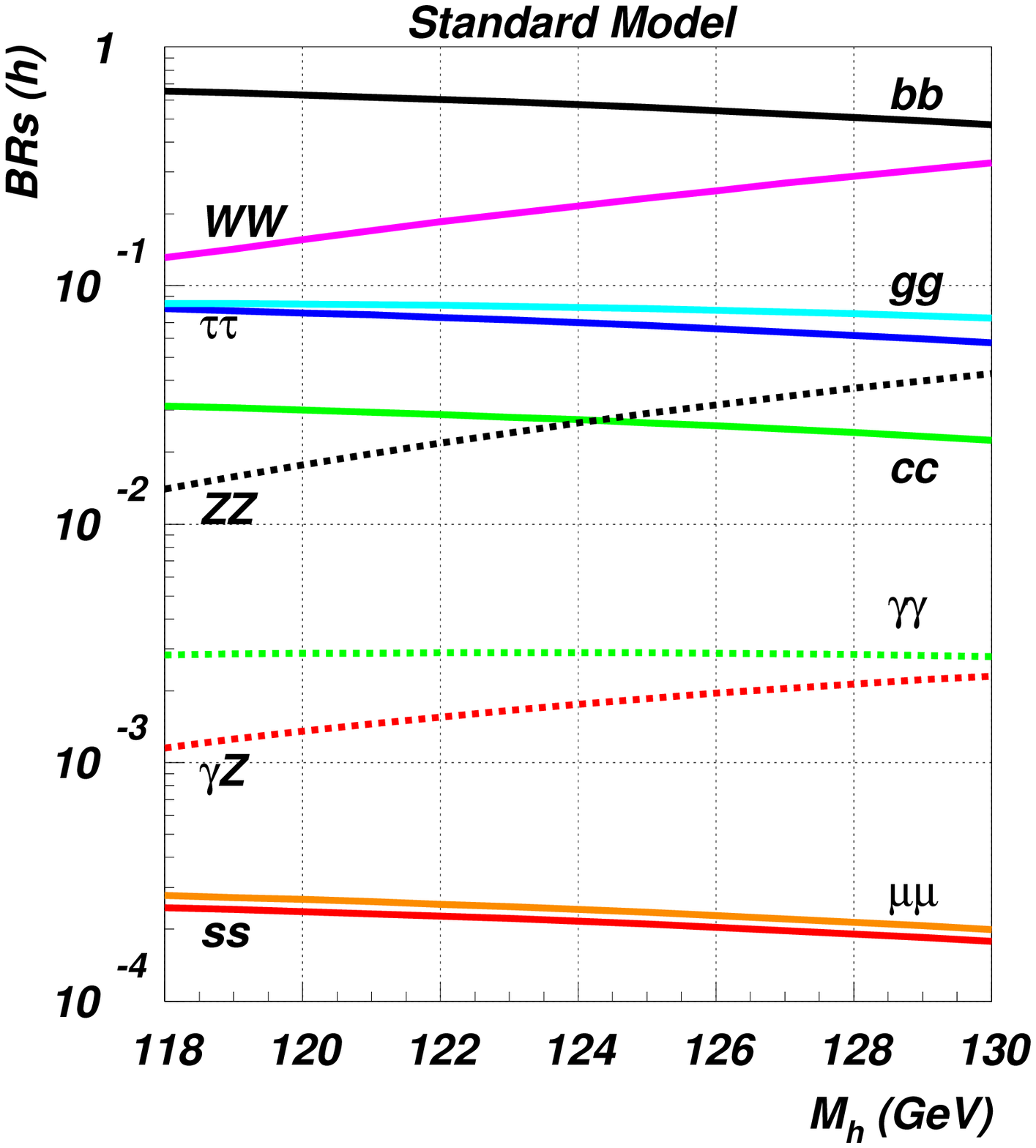}}
 \\
  \subfloat[\model, $\Pa$]{ 
  \label{BRH1_tb1_c1}
  \includegraphics[angle=0,width=0.48\textwidth ]{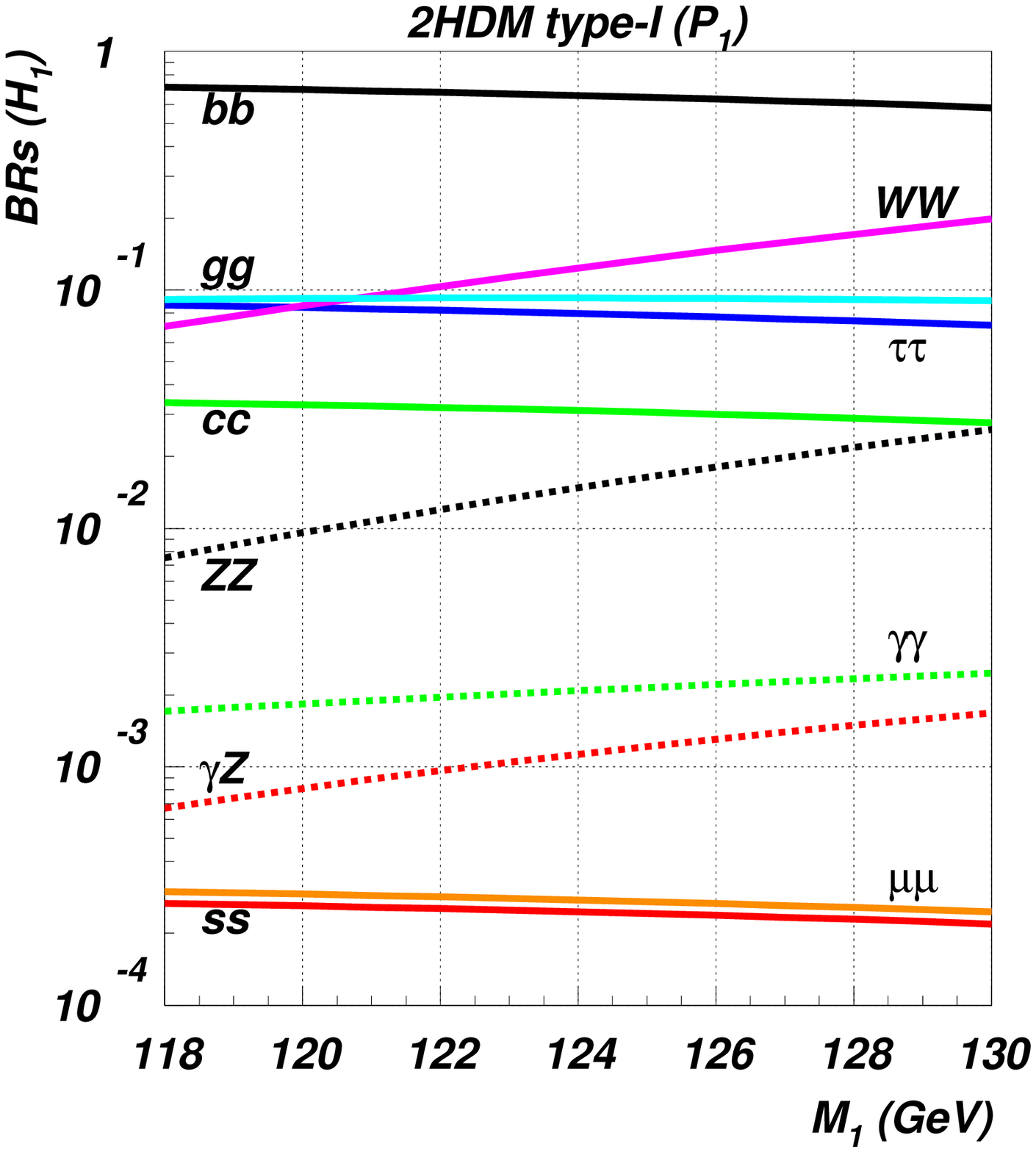}}
  \subfloat[\model, $\Pb$]{
  \label{BRH1_tb1_c2}
  \includegraphics[angle=0,width=0.48\textwidth ]{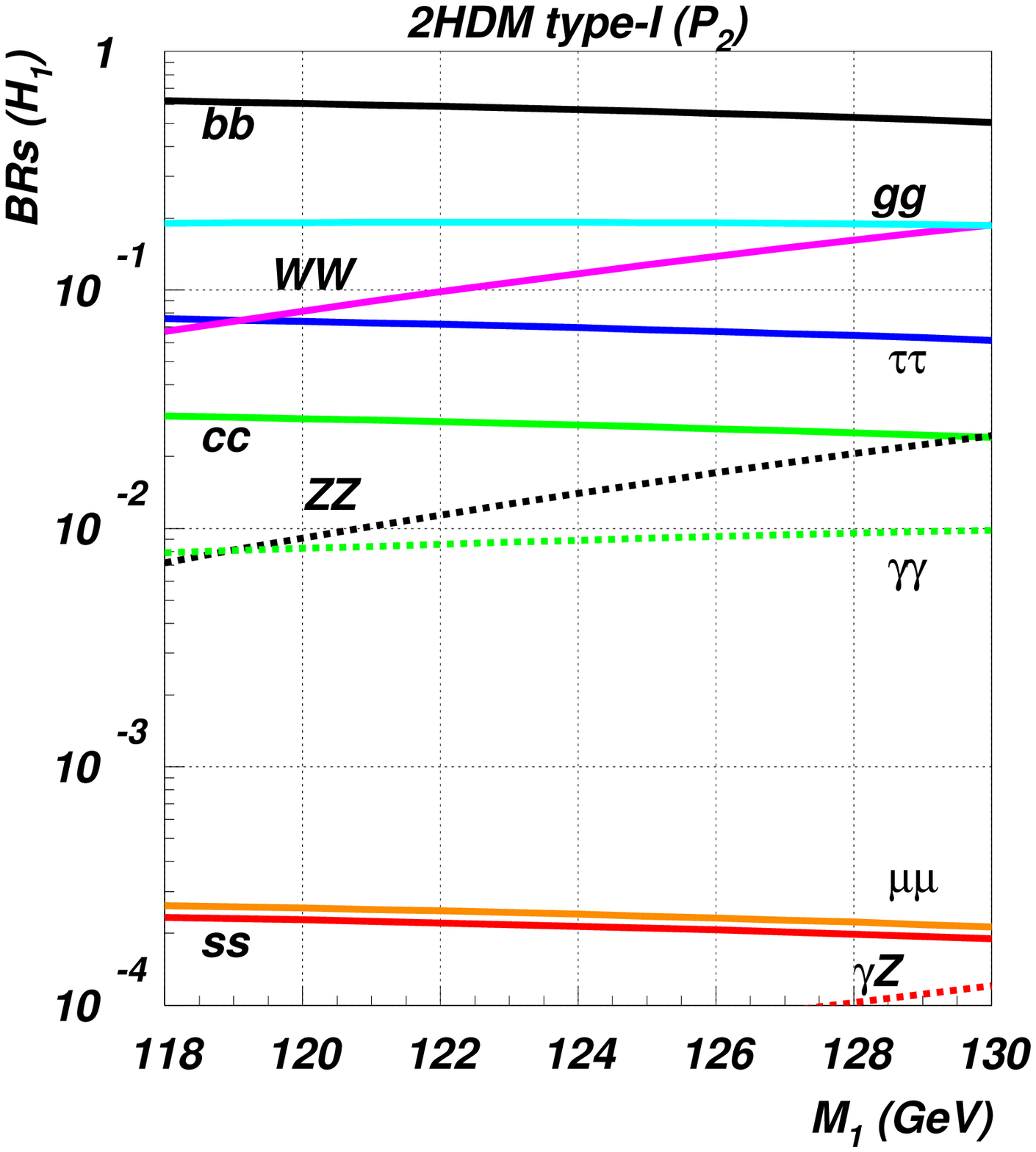}} 
\caption{Branching ratios of the (lightest) neutral Higgs
  plotted against its mass.
  The SM case is shown in panel~\ref{BRH1_tb1}.
  Panels~\ref{BRH1_tb1_c1} and \ref{BRH1_tb1_c2} are respectively for
  the benchmark points
  $\Pa$ and $\Pb$ 
  in the \model.
  The parameters are set as in~(\ref{eq:default}) unless stated otherwise.}
\label{fig:BRH1}
\end{figure}
As for $H_1$,
we plot its branching fractions in the vicinity of $M_1 = 125\GeV$,
in figures~\ref{fig:BRH1}.
One finds that the $b\overline{b}$ mode is dominant
at both benchmark points.
This tends to be the case unless
both $\sa{1}$ and $\sa{2}$ are vanishingly small.
At the point where $\sa{1} = \sa{2} = 0$,
the tree-level $H_1$--fermion coupling vanishes,
causing its fermiophobia.
Considering these dominant decay modes of $\Hc$ and $H_1$,
we are going to concentrate on the final state,
$\tau\nu b\overline{b}$ for numerical simulation.

One can also notice that the branching fractions of
the $\gamma\gamma$ and $gg$ modes can differ significantly
from their individual SM values.
Albeit not directly related to the charged Higgs search,
these changes greatly affect the phenomenology of $H_1$.
We shall come back to this issue in section~\ref{sec:H1}.

\section{LHC prospects of \boldmath$\Hc$ search}
\label{sec:res}


The main process of our concern is
$pp \rightarrow W \rightarrow \Hc H_1$, followed by the decays,
$\Hc\rightarrow \tau\nu$ and
$H_1\rightarrow b\overline{b}$ at the LHC\@.
Among these three subprocesses,
the charged Higgs production and decay revealed their outstanding
stability under variation of $\tb$ over a wide range,
in the preceding section.

We shall envisage two experimental conditions at the LHC,
which we call the ``early'' and the ``full'' searches.
The ``early search'' denotes the centre-of-mass energy of
the current run, $\rs = 7\TeV$, and the integrated luminosity of
$L = 10\ifb$.
We assume that the ``full search'' will deliver $L = 100\ifb$
at $\rs = 14\TeV$.

The kinematics of the $\tau\nu b\overline{b}$
state can be fully reconstructed with
high efficiency as it has only one neutrino.
We assume a
$\tau$-reconstruction efficiency $\sim 100\%$ and a
$b$-tagging efficiency $\sim 70\%$,
plus the following set of selection cuts:
\begin{equation}
  \label{eq:cuts}
  \begin{split}
p_T(\bbbar), p_T(\tau^\pm) &>25\GeV, \\
|\eta (\bbbar)|, |\eta (\tau^\pm)| &<2.5, \\
p_T(\parbar{\nu})&>20 \GeV, \\
|\eta (\parbar{\nu})|&<4.5, \\
M(b, \bar{b}) &= M_1 \pm 10\GeV ,
  \end{split}
\end{equation}
which we choose considering the operation of ATLAS\@.
Assuming these cuts, we calculated the SM
background.\footnote{Including contributions from a
  SM Higgs with mass of $125$ GeV.}
The cross-section is
$\sigma(pp \to \tau^\pm\nu b\overline{b})_\mathrm{BG}\simeq 55\fb$ at
$\rs = 7\TeV$ and $\simeq 110\fb$ at $\rs = 14\TeV$.

Along with the associated production of $\Hc H_1$,
we also study a similar process with $\Hc$ replaced by $W$, i.e.\
$pp\rightarrow W \rightarrow W H_1$.
As we shall see, the parameter space covered by
this Higgs-strahlung is complementary to the region
accessible to the $\Hc H_1$ state.
We select the final state, $l^\pm\nu b\overline{b}$ with $l = e,\mu$,
and assume a $W$-reconstruction efficiency $\sim 100\%$.
The cuts in~(\ref{eq:cuts}) are used again, and
the same cut on $\tau$ is applied on the transverse momenta
of the light leptons, $p_T(l^\pm) > 25\GeV$.
Using these parameters, we obtain the Higgs-less SM background
cross-section to be
$\sigma(pp \rightarrow l^\pm\nu b\overline{b})_\mathrm{BG}\simeq 70\fb$
at $\rs = 7\TeV$ and $\simeq 140\fb$ at $\rs = 14\TeV$.

We used LanHEP \cite{Semenov:1996es} to implement the model,
and CalcHEP \cite{calchep} 
plus the CTEQ6M parton distribution functions
\cite{Nadolsky:2008zw},
to perform the phenomenological analysis.
The Passarino-Veltman functions appearing in the effective vertices
of $H_i$--$g$--$g$ and $H_i$--$\gamma$--$\gamma$ are calculated
by calling the LoopTools package \cite{Hahn:1998yk}.

We first check the rates of our main processes
with $\sa{2} = \sa{3} = 0$.
In this $CP$-conserving case, one has the lighter and the heavier
$CP$-even Higgses, $h\equiv H_1$ and $H\equiv H_2$,
and the $CP$-odd Higgs, $A\equiv H_3$.
\begin{figure}
  \centering
  \subfloat[$\MHc = 86\GeV$]{
    \label{fig:CPC xs vs angles}
  \includegraphics[width=0.48\textwidth]{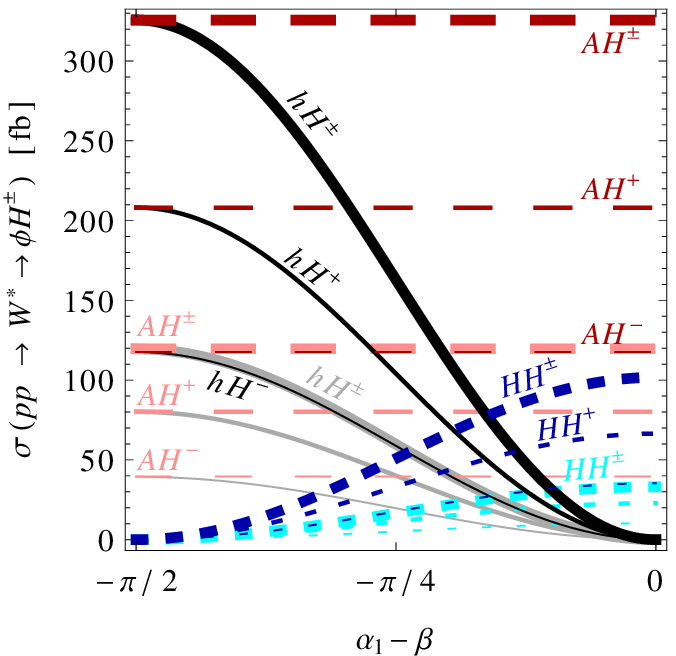}}
  \quad
  \subfloat[$\max_{\alpha_1 - \beta} \sigma$]{
    \label{fig:CPC xs vs MHc}
  \includegraphics[width=0.48\textwidth]{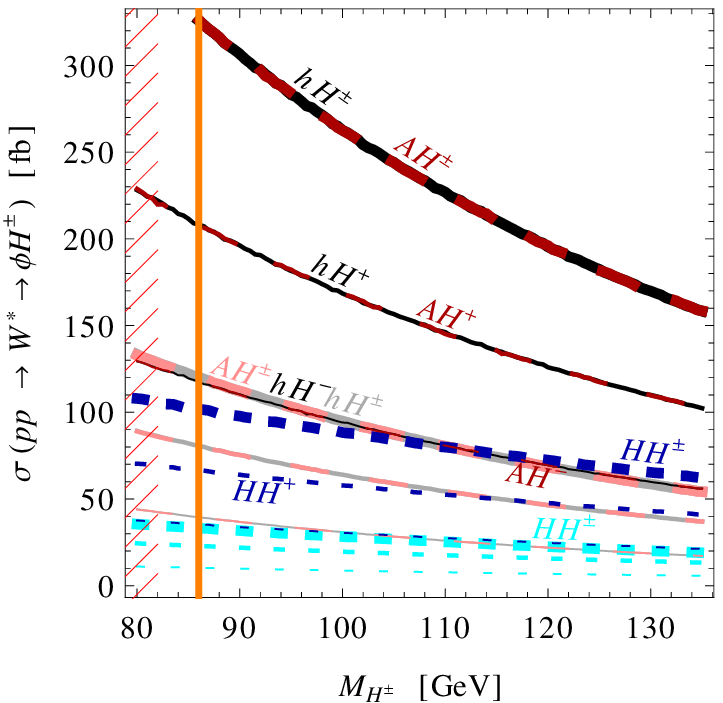}}
  \caption{Production cross-section of each neutral Higgs in association with
    a charged Higgs via a virtual $W$-boson
    in the $CP$-conserving case.
    The masses are $M_h = M_A = 125\GeV, M_H = 200\GeV$.
    For each channel, the lighter and the darker curves represent
    $\rs = 7\TeV,14\TeV$, respectively.
    In panel~\ref{fig:CPC xs vs MHc},
    the hatched region is excluded by LEP at 95\% CL \cite{Abbiendi:2008aa},
    and each curve is for
    $\alpha_1 - \beta$ that maximises the cross-section.}
  \label{fig:CPC xs}
\end{figure}
The production cross-sections of $h$, $H$, and $A$,
each in association with a charged Higgs,
are shown in figure~\ref{fig:CPC xs vs angles}
as functions of $\alpha_1 - \beta$.
For each type of neutral Higgs,
there are three curves that respectively
represent $H^-$, $H^+$, and $\Hc$ productions
in the order of increasing thickness.
The higher $H^+$ rate relative to $H^-$ reflects the fact that the LHC
is a $pp$ collider, not $p\overline{p}$.
The shape of each curve is essentially determined by
the dependence of the corresponding $W$--$\Hc$--$H_i$ coupling
on $\alpha_1$ and $\beta$, presented in table~\ref{tab:feynman rules}.
These patterns
will be useful for understanding the simulation results to follow.
 From the plot, one can expect $\order(10^4)$ events that occur through
a charged-neutral Higgs associated production
with the ``full luminosity'' at $14\TeV$, unless
the mixing angle is such that the given channel is highly suppressed.

Our analysis is of relevance to a light charged
Higgs, even if one is not interested in a scenario where charged
Higgs bosons have been on-shell produced at LEP energies.
In this regard, it is legitimate to consider
changes that would be caused by varying $\MHc$.
Obviously, the above production rates will be reduced as $\MHc$ grows,
as is shown in figure~\ref{fig:CPC xs vs MHc}.
We take the range of $\MHc$ up to $\sim 135\GeV$,
beyond which the changed Higgs decay pattern undergoes
a qualitative change \cite{Ginzburg:2012hc},
thereby invalidating our assumptions given in~(\ref{eq:BRs of H+}).
For $\MHc \sim 135\GeV$,
each cross-section decreases roughly down to half the value
in panel~\ref{fig:CPC xs vs angles}.
These curves illustrate the phase-space
suppression that takes place for higher $\MHc$.
We emphasise that this reduction of statistics is the only
essential information that one needs to understand
how the significances would change
in the following quantitative analysis with $\MHc = 86\GeV$,
if one took a different $\MHc$.


For the simulation of charged Higgs search,
we turn on $CP$-violation allowing for mixing among all the three
neutral Higgses, $h$, $H$, and $A$.
\begin{figure}
  \subfloat[$\rs=7\TeV, L=10\ifb$]{ 
  \label{overlap_07TeV}
  \includegraphics[angle=0,width=0.48\textwidth ]{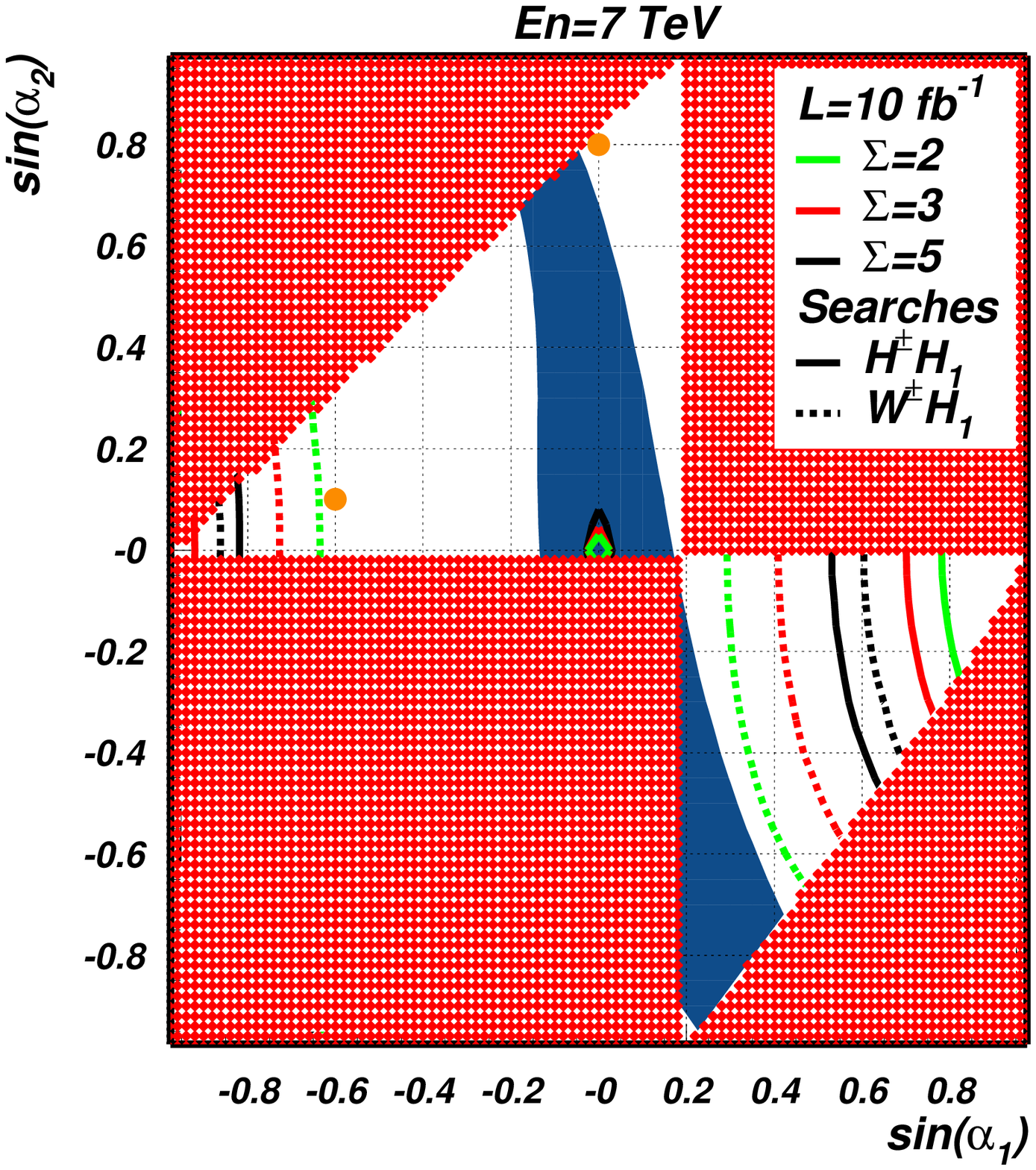}}
  \subfloat[$\rs=14\TeV, L=100\ifb$]{
  \label{overlap_14TeV}
  \includegraphics[angle=0,width=0.48\textwidth ]{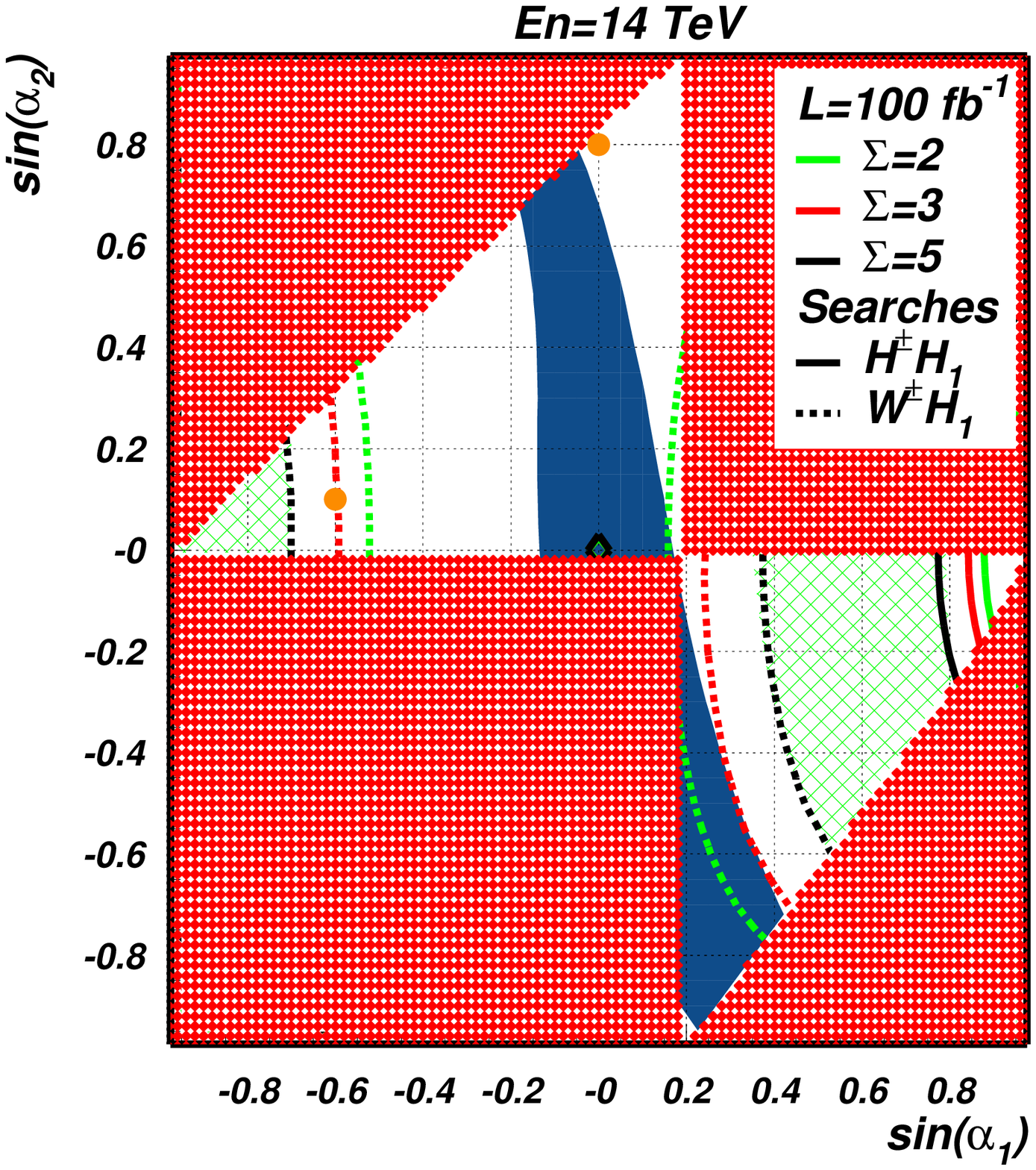}}  
\caption{Significance contours
for the processes,
$pp\rightarrow H^\pm H_1\rightarrow \tau^\pm \nunubarl{\tau} b \bar{b}$
(solid lines)
and
$pp\rightarrow W^\pm H_1\rightarrow l^\pm \nunubarl{l} b \bar{b}$
(dashed lines),
on the
$(\sa{1},\sa{2})$ plane in the $CP$-violating case.
The parameters are set as in~(\ref{eq:default}) unless stated otherwise.
Light grey (green online), dark grey (red online),
and black correspond to the significances $\Sigma = 2,3,5$, respectively.
The medium-shadowed regions (red online) are excluded by theoretical constraints.
The dark-shadowed region (blue online) is ruled out by
current limits on $H_{2,3}$.
The light-shadowed region (green online) allows for
discovery of both processes.}
\label{fig:significance CPV}
\end{figure}
In figures~\ref{fig:significance CPV},
the contours of the significance $\Sigma$ are shown for each process,
which have been evaluated by means of
the method described in~\cite{def significance}.
There are two complications in comparison to the $CP$-conserving limit.
One is that the additional mixing between the
$CP$-odd and the $CP$-even components
affects the significance of each process.
The other is that $M^2_{1,2,3}$ are no longer independent of one another
but constrained by~(\ref{eq:M3}).
In combination with the latter relation,
the mass ordering~(\ref{eq:mass ordering})
excludes the upper-right and the lower-left rectangular regions
from the parameter domain.
Another theoretical constraint that rules out
the upper-left and the lower-right triangular regions, is
perturbativity~(\ref{eq:perturbativity}).
We have also checked numerically that imposing perturbativity
guarantees the tree-level unitarity in Higgs-Higgs scattering
in the parameter space under consideration.

In addition to the above theoretical requirements,
we impose the experimental constraints in~(\ref{eq:H2 exclusion})
in order to respect the current Higgs production limits from the LHC\@.
This rules out the dark-shadowed strip on the plane.
The bounds on $H_2$ and $H_3$ apply as long as their masses are
in the range between 160 and 600~GeV\@.
All these conditions being fulfilled,
the question to ask is how much portion of the
remaining unshaded region the LHC can cover.

As in the preceding plot,
the $W$--$\Hc$--$H_1$ coupling largely determines
the shape of the $\Hc H_1$ contours.
As long as one moves along the horizontal axis,
$H_1$ is purely $CP$-even, and therefore its interactions
not involving other neutral Higgses
reduce to those of $h$ in the $CP$-conserving Higgs sector.\footnote{
  Nevertheless, the interactions of $H_2$ and $H_3$ do not reduce to
  those of $H$ and $A$ due to non-vanishing $\sa{3}$.}
For instance, the $W$--$\Hc$--$H_1$ coupling vanishes at the point
$(\sa{1}, \sa{2}) = (\sin\!\beta,0) = (0.98,0)$,
as is the case in figure~\ref{fig:CPC xs vs angles}.
This explains why the significance decreases
as one approaches this point.

It should be meaningful to interpret the interesting behaviour of $\Sigma$
around the point $\alpha_1 = \alpha_2 = 0$,
although it belongs to the $H_{2,3}$ exclusion strip.
For instance, there might be a way to circumvent
the constraint~(\ref{eq:H2 exclusion}) by taking
a different set of parameters.
In the neighbourhood of this point,
one finds a sudden drop of significance
even though the $W$--$\Hc$--$H_1$
coupling stays nearly maximal.  
This is because the $H_1$--$b$--$\bar{b}$ coupling is suppressed there
as can be seen in table~\ref{tab:feynman rules}.
At the origin, $H_1$ becomes fermiophobic,
and its dominant decay modes are $W W^*$ and $Z Z^*$.
The fermiophobia of $H_1$ generically gives rise to an enhanced
$\BR(H_1\to\gamma\gamma) \sim 1\%$
without the charged Higgs loop contribution.
Including the $\Hc$-loop, the diphoton branching fraction
can range from zero up to $\order(10\%)$.
[We give a related discussion in section~\ref{sec:H1}.]
Therefore, the possibility of observing the $\Hc H_1$ state
in the central low-significance hole
via an alternative decay channel of $H_1$
such as the $\gamma\gamma$ mode,
is highly dependent on the other model parameters.

As one departs away from the horizontal axis, $\sa{2} = 0$,
$H_1$ acquires more of the $CP$-odd component.
This reinforces the production rate of $\Hc H_1$ and thereby the significance,
since the $W$--$\Hc$--$A$ coupling is not suppressed for any $\alpha_1$
as can be seen in figure~\ref{fig:CPC xs vs angles}.

Overall, figures~\ref{fig:significance CPV} demonstrate that
there is an ample portion of the parameter space in which
the LHC has a good chance to discover the $\Hc H_1$ production.
Obviously, the chance is substantially better in the ``full luminosity'' case,
shown in figure~\ref{overlap_14TeV}.
With the exception of the narrow corners
around both ends of the horizontal axis
and the hole encircling the fermiophobic point, 
the discovery of $\Hc H_1$ through the $\tau\nu b\overline{b}$ final state
should be possible at any point
in the unshaded physically sensible domain
defined by~(\ref{eq:mass ordering}), (\ref{eq:perturbativity}), and
(\ref{eq:H2 exclusion}).

In the same figures is presented the $H_1$-strahlung process,
which could also be detected depending on the mixing angles and the
LHC operation condition.
One can understand the shape of the $W^\pm H_1$ contours
by looking at the $H_1$--$W^+$--$W^-$ coupling
in table~\ref{tab:feynman rules}.
In figure~\ref{overlap_07TeV} for ``early searches'',
one finds no common portion of the parameter space
with $\Sigma>5$ for both processes.
This setup predicts that
it is not possible to discover both channels during the first data
taking period in the $l\nu b\bar{b}$ final states.
Should this occur, this scenario will have to be modified to survive.
In figure~\ref{overlap_14TeV}, one finds a much broader region
in which $H_1$-strahlung could be discovered.
Part of this region is shadowed in light grey where
the ``full LHC'' could observe
both this process and $\Hc H_1$ production.



\section{\boldmath Impact of $\Hc$-loop on
  $H_1\rightarrow\gamma\gamma$}
\label{sec:H1}

This section is devoted to phenomenology of the lightest neutral Higgs.
In particular, we show that
the charged Higgs loop can exert a marked influence on
the $H_1\to\gamma\gamma$ decay mode,
which is of the utmost importance in the neutral Higgs search
at the LHC\@.
We shall consider the following standard Higgs production mechanisms:
\begin{align*} 
gg &\rightarrow H_1 &
&\text{(gluon-gluon fusion),} \\
qq &\rightarrow jj + H_1 &
&\text{(vector boson fusion),} \\
q\overline{q} &\rightarrow W/Z + H_1 &
&\text{(Higgs-strahlung from $W/Z$),} \\
gg,q\overline{q} &\rightarrow t\overline{t} + H_1 &
&\text{(associated production with a top-quark pair).}
\end{align*}
In figures~\ref{fig:np},
we plot the cross-sections of these channels at $\rs = 7\TeV$
against $\tb$ for each of the two benchmark points.
\begin{figure}
  \subfloat[$\Pa$]{ 
  \label{np_tb1_c1_07}
  \includegraphics[angle=0,width=0.48\textwidth ]{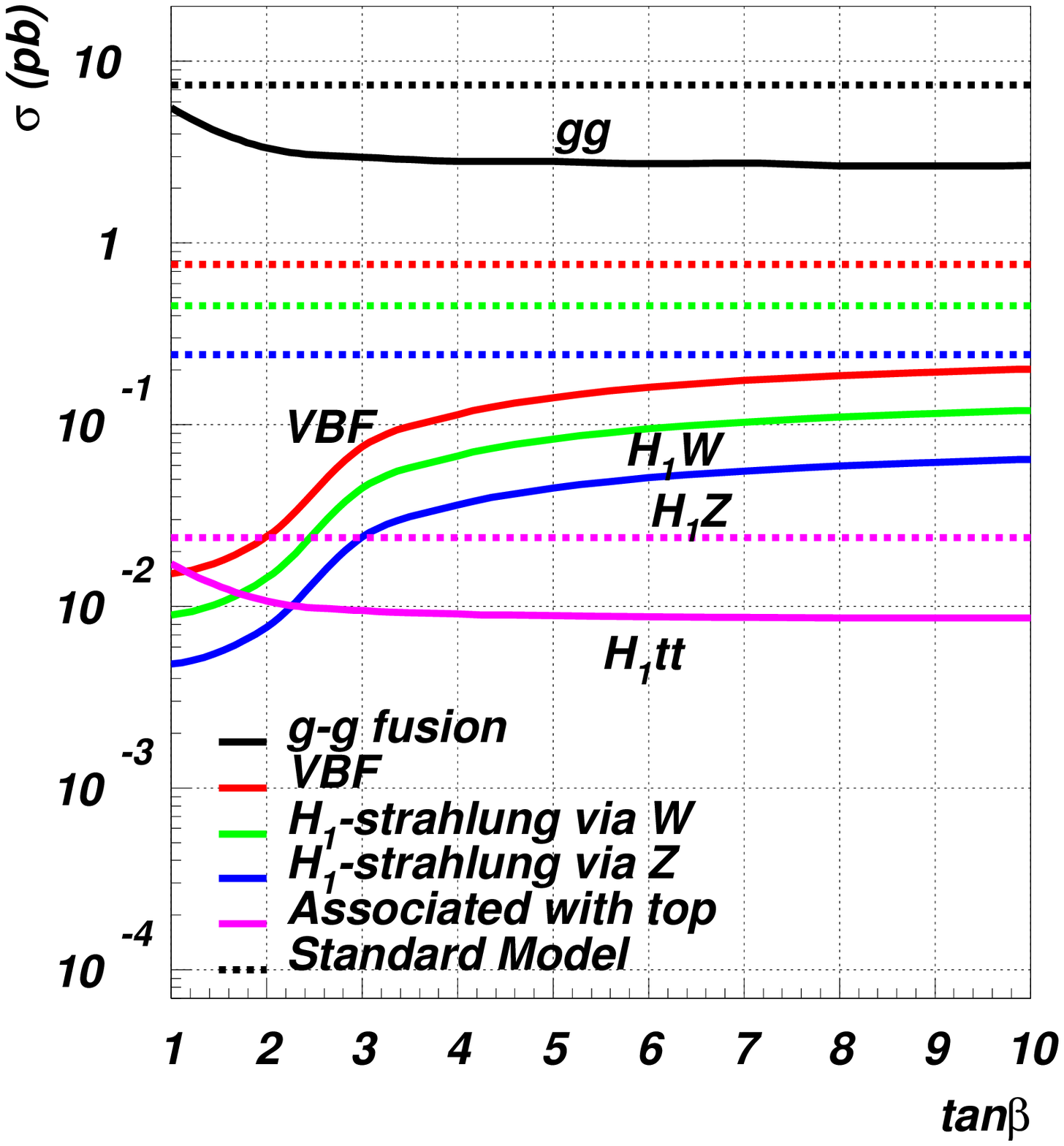}}
  \subfloat[$\Pb$]{
  \label{np_tb1_c2_07}
  \includegraphics[angle=0,width=0.48\textwidth ]{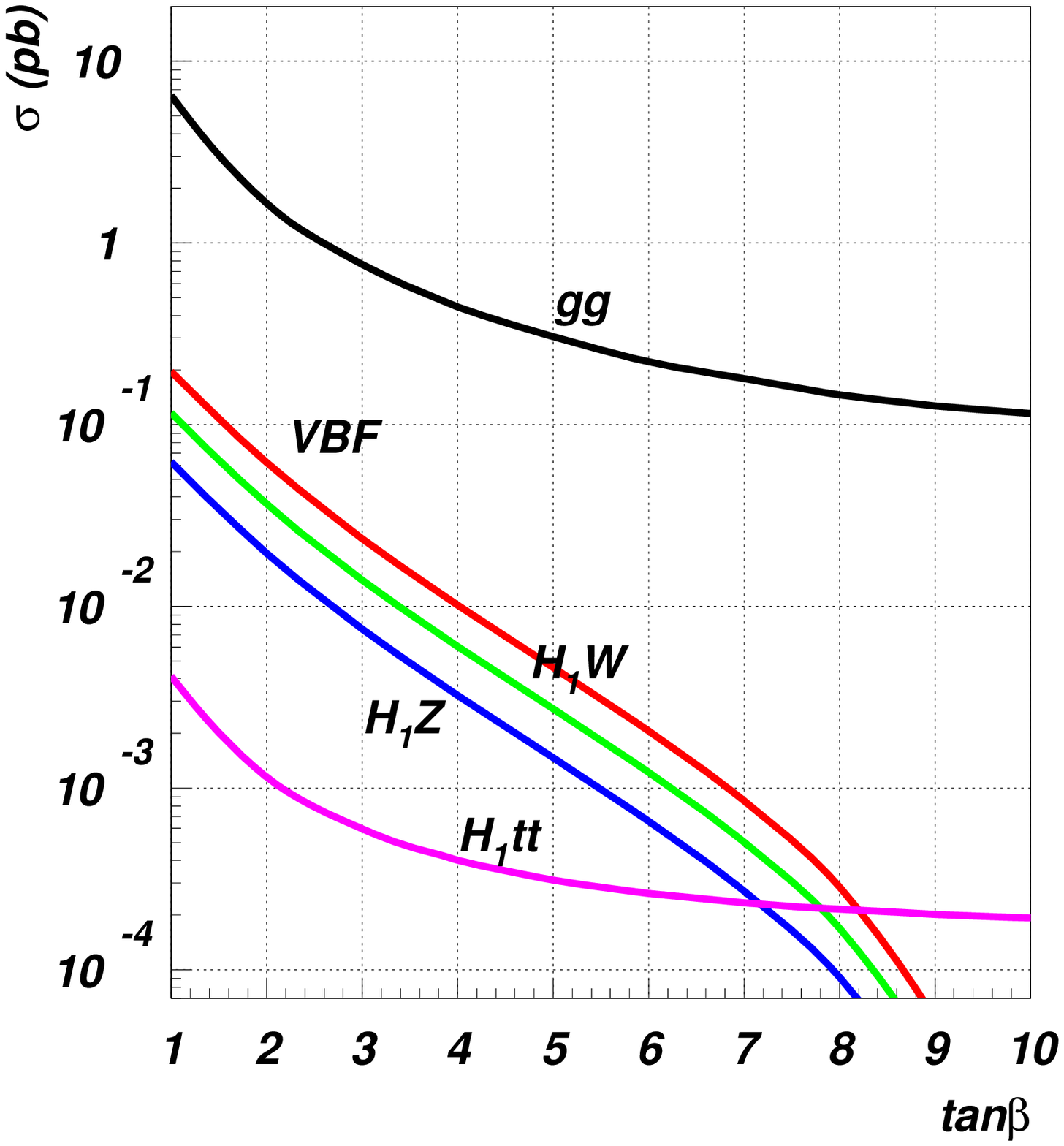}} 
\caption{Cross-sections of the lightest neutral Higgs production mechanisms plotted against $\tb$ at $\rs = 7\TeV$
for the benchmark points $\Pa$ and $\Pb$.
The parameters are set as in~(\ref{eq:default}) unless stated otherwise.
The curves are independent of $\mu$.}
\label{fig:np}
\end{figure}
We have explicitly verified that all these production mechanisms are
mostly unaffected in the range of $M_1$ between $118$ and $130\GeV$
and in the neighbourhood of the chosen $\MHc$.
All the production rates are lower than those in the SM\@.
This is because of the suppression of the couplings
in table~\ref{tab:feynman rules}.
Combining production and decay, we obtain an intriguing result:
$H_1$ may have escaped from the past and present searches due to
the suppression of both its production and the branching ratios
to massive vector boson pairs.
As this scenario generically predicts lower
rates into fermion pairs,
it could be disfavoured by the recent hint in the $b\overline{b}$ mode
from the Tevatron \cite{TEVNPH:2012ab} if it persists.
Obviously, one can ease the reduction of $H_1$ production
by taking the Higgs mixing pattern to the SM-like limit,
$(\sa{1},\sa{2}) = (1,0)$.
In this case, the charged-neutral Higgs associated production
will be suppressed as shown in figures~\ref{fig:significance CPV}.
Recall that we are assuming a hierarchical neutral Higgs mass spectrum
and that all the experimental evidences
from the LHC and the Tevatron are caused by the lightest of
the three neutral Higgs particles, which is also supposed to
be produced in association with $\Hc$.
Relaxing one or more of these hypotheses may lead to
a different consequence, although we do not explore this possibility
in this paper.

On the other hand,
branching fractions of the two loop-induced decay modes,
$H_1 \rightarrow \gamma\gamma$ and $H_1 \rightarrow gg$,
have possibilities of being enhanced as well as diminished,
depending on the input parameters.
If they are increased, the suppressed Higgs production can be alleviated.

The decay into the digluon final state arises from the fermion loop,
dominated by the top-quark contribution.
Although the same type of loop is present also in the SM,
its value can be different due to the variation of
the $H_1$--$t$--$\overline{t}$ coupling
which is determined by the Higgs mixing angles
as in table~\ref{tab:feynman rules}.
In particular, an interesting role is played by the $CP$-odd
component $A$ in $H_1$ that could enhance the digluon branching fraction.
Its effects are twofold:
(a) $A$ does not couple to a vector boson pair unlike
the $CP$-even Higgs $h$;
(b) the digluon to difermion partial width ratio
of $A$ is higher than the corresponding quantity of $h$, i.e.\
\begin{equation}
\label{eq:Agg/hgg}
  \frac{\Gamma (A \rightarrow gg)}
  {\Gamma (A \rightarrow f\overline{f})} = 2.3\ 
  \frac{\Gamma (h \rightarrow gg)}
  {\Gamma (h \rightarrow f\overline{f})} ,
\end{equation}
when $M_A = M_h = 125\GeV$.
This effect stems from the different Lorentz structures
of the scalar and the pseudo-scalar couplings
with fermions.  
The potential enhancement of the $gg$ decay mode might be of interest
at a future lepton collider experiment.

The decay amplitude of $H_1 \rightarrow \gamma\gamma$
has two more components, the $W$- and the $\Hc$-loops,
in addition to the fermion loop.
In the SM, the dominant contribution comes from the $W$-loop.
Like the fermion loop, the $W$-loop can vary as a function of $R_{ij}$.

One might attempt to increase the diphoton branching fraction
by invoking the same mechanism as with the digluon mode.
As $H_1$ becomes pseudo-scalar-like,
the fermion loop alone does receive the relative enhancement
that is given by~(\ref{eq:Agg/hgg})
with $gg$ replaced by $\gamma\gamma$ and each numerator
including only the fermion loop.
However, this kills the $W$-loop at the same time,
which is the dominant contribution in the SM\@.
Therefore, a high $CP$-odd fraction in $H_1$ is not really
helpful in increasing $\BR(H_1\rightarrow\gamma\gamma)$.

Nevertheless, it is known that
the charged Higgs loop can greatly affect the $\gamma\gamma$ mode
when $\MHc$ is low
\cite{gamma-gamma fusion,Posch:2010hx,Arhrib:2012ia}.
In particular, this mechanism can lead to a high enough diphoton rate
to be consistent with the excess surrounding the Higgs mass of $125\GeV$
recently observed by ATLAS \cite{ATLAS:2012ad}
and CMS \cite{Chatrchyan:2012tw}.
The $H_1$--$H^+$--$H^-$ coupling
depends on more parameters than the other loops, in particular on $\mu$,
with which one can play in order to engineer the amplitude.
The maximal permitted size of the $\Hc$-contribution
is mostly controlled by perturbativity and unitarity.
In passing, we remark that change of $\mu$ affects also
$\Gamma(H_{2,3}\to H^+ H^-)$ and therefore
the region excluded by~(\ref{eq:H2 exclusion}),
as we will see in the following plots.


\begin{figure}
  \subfloat[Branching ratio, $\mu=100\GeV$]{ 
  \label{H1-AA}
  \includegraphics[angle=0,width=0.48\textwidth]{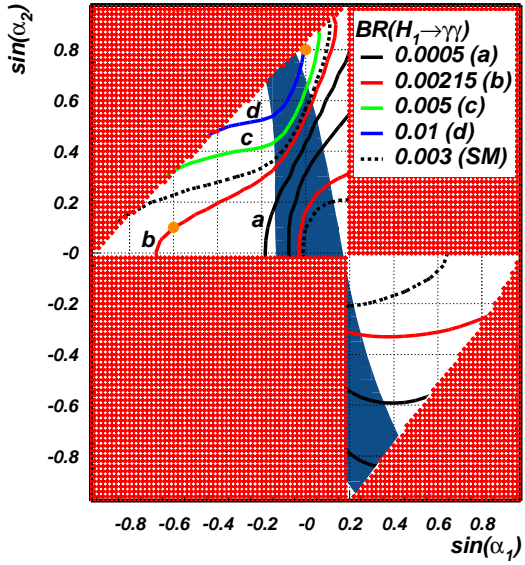}}
  \subfloat[Cross-section, $\mu=100\GeV$]{ 
  \label{gg-AA}
  \includegraphics[angle=0,width=0.48\textwidth]{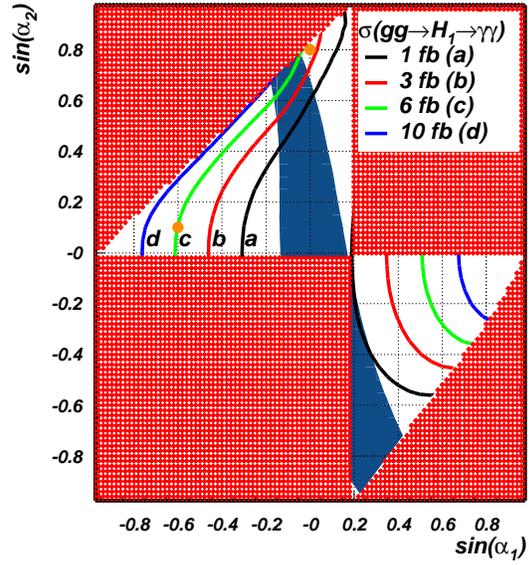}}
\\
  \subfloat[Branching ratio, $\mu=180\GeV$]{ 
  \label{H1-AA mu=180}
  \includegraphics[angle=0,width=0.48\textwidth]{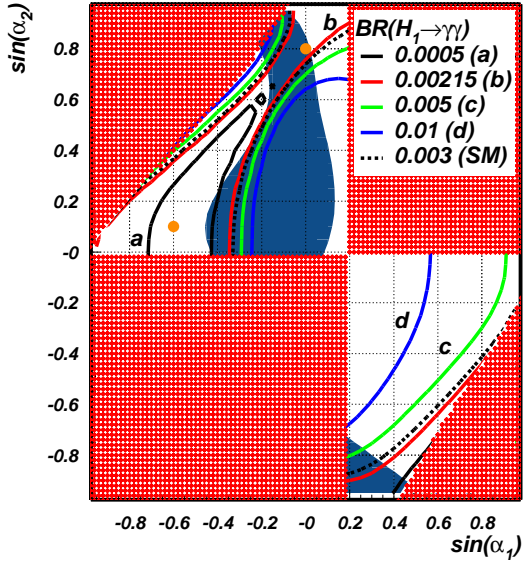}}
  \subfloat[Cross-section, $\mu=180\GeV$]{ 
  \label{gg-AA mu=180}
  \includegraphics[angle=0,width=0.48\textwidth]{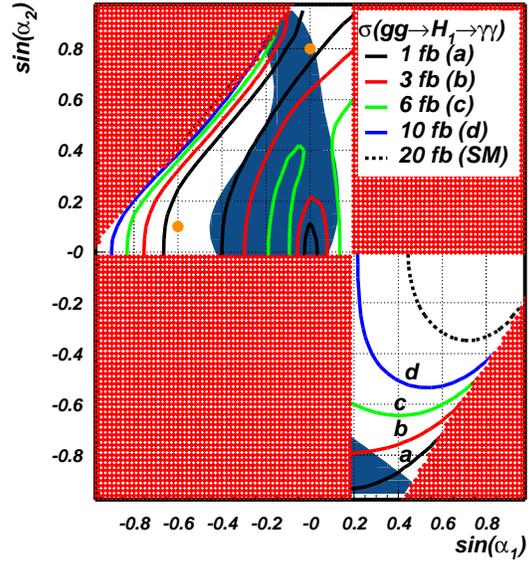}}
\caption{Contours of the branching ratios and the cross-sections involving the diphoton final state on the $(\sa{1}$, $\sa{2})$ plane at $\rs=7\TeV$.
The two benchmark points $P_1$ and $P_2$ are marked by dots.
The parameters are set as in~(\ref{eq:default}) unless stated otherwise.
The medium-shadowed regions (red online) are excluded by theoretical constraints.
The dark-shadowed region (blue online) is ruled out by
current limits on $H_{2,3}$.}
\label{fig:ggH1}
\end{figure}
To further study this potential variation,
we plot the branching fraction of $H_1 \rightarrow \gamma\gamma$
in figures~\ref{H1-AA} and \ref{H1-AA mu=180}.
Indeed, it can deviate substantially from its SM value
depending on the mixing angles.
One finds a valley between the two curves labelled ``a''
on each of the two plots,
along which the branching ratio is highly suppressed.
Across this valley, the $H_1 \rightarrow \gamma\gamma$ amplitude
flips its sign.  
As the plots show, the contours are deformed if $\mu$ is changed,
since they depend on the $\Hc$-loop.
One can also notice that part of the ``d'' contour of a high
branching fraction lies on the verge of the perturbativity border,
demonstrating that a sizeable $\Hc$-loop contribution involves
large quartic Higgs couplings.

In conjunction with gluon-gluon fusion, this brings us to
the cross-section of $gg \rightarrow H_1 \rightarrow \gamma\gamma$
in figures~\ref{gg-AA} and \ref{gg-AA mu=180}.
With our crude leading-order approximation,
it is estimated to be about $20\fb$ in the SM\@.
In the entire region of panel~\ref{gg-AA}, the cross-section is smaller than
its SM value due to the suppressed $H_1$ production.
Nevertheless, one still has the option to change $\mu$
that affects the charged Higgs loop.
For instance, the cross-section at $(\sa{1},\sa{2}) = (0.5,0)$
is approximately $6\fb$ for $\mu = 100\GeV$.
At this point, the $W$- and the $\Hc$-loops interfere destructively.
%
Taking $\mu = 180\GeV$ instead, 
one can make the interference constructive by flipping the sign
of the $\Hc$-contribution.
This pushes the cross-section up above the SM prediction,
as displayed in figure~\ref{gg-AA mu=180}.
In this plot,
the rate of $gg \rightarrow H_1 \rightarrow \gamma\gamma$
can be comparable to its SM value even when
the mixing angles are
fairly removed from the SM-like point, $(\sa{1},\sa{2}) = (1,0)$.
In combination with figures~\ref{fig:significance CPV},
this exhibits the exciting possibility that
the LHC can discover a light charged Higgs in the foreseeable future
while the recent evidences for Higgs decays to $\gamma\gamma$ are partly
due to the charged Higgs loop effect.

As one can expect, the charged Higgs loop effect weakens for high $\MHc$.
For instance, $\BR(H_1\rightarrow\gamma\gamma)$ for $\MHc \sim 135\GeV$
is roughly half its value for $\MHc = 86\GeV$,
when one keeps $(\sa{1},\sa{2}) = (0.5,0)$ and $\mu = 180\GeV$.

\section{Conclusions}
\label{sec:conclusions}

We demonstrated the superb potential of the LHC to discover
a light charged Higgs close to the $W$-boson mass
in the $CP$-violating \model\@.
For this purpose, we exploited the property of the
$pp\rightarrow W\rightarrow H_1 \Hc$
process that its cross-section is stable against variation of $\tb$,
which can be anywhere above the lower bound
placed by the most stringent FCNC constraints.
We set the Higgs masses to their favoured values,
$\MHc \sim 86\GeV$ and $M_1 \sim 125\GeV$, and
found that the LHC can spot an excess in the
$\tau\nu b\overline{b}$ events,
except in small corners and holes of the parameter space.
We also examined the $H_1$-strahlung channel from $W$,
which offers access
to the complementary parameter volume to that of $H_1 \Hc$.

With regard to $CP$-violation, we observed interesting consequences of
the scalar--pseudo-scalar mixing.
In particular, $\BR(H_1\to gg)$ can
be considerably affected by the composition of $H_1$.
This result might be relevant to Higgs scrutiny at
prospective linear collider experiments.

Motivated by the recent evidences for Higgs production at the LHC,
we studied
$\sigma(gg\rightarrow H_1\rightarrow \gamma\gamma)$.
Due to the Higgs mixing matrix elements appearing at
interaction vertices, the production of $H_1$ is generically suppressed
in comparison to the SM\@.
Nonetheless, the $\Hc$-loop amplitude can make
a constructive interference with the standard $W$-loop
to raise $\BR(H_1\to\gamma\gamma)$.
Depending on the parameter choice,
this effect can be significant enough to increase
the diphoton production rate from gluon-gluon fusion
up to or even above the SM prediction.
Alternatively, one can opt to enhance the $H_1$ production itself
by letting the Higgs mixing pattern approach the SM-like limit.
In this case, the other processes with $H_1$ going into
the fermion and the massive vector boson pairs are also restored,
the price to pay being suppression of the $H_1 \Hc$ rate.

Related to this last point,
it may be beneficial to study further possibilities
of the light charged Higgs scenario in which
one makes assumptions different from those underlying
the present article.

\begin{acknowledgments}
GMP would like to thank Per Osland and Alexander Pukhov for helpful discussions.
JP thanks Eung Jin Chun for valuable comments and
Kazuki Sakurai for the help with access to the literature.
We acknowledge financial support from
German Research Foundation DFG through Grant No.\ STO876/2--1
and BMBF\@.
\end{acknowledgments}

\appendix


\section{Gauge-fixing Lagrangian of the \thdm} 
\label{appe:b}
Assuming the 't Hooft-Feynman gauge, we choose a basis for the $\Phi_i$ by expanding the Higgs doublets as
\begin{equation}\label{Higgs_goldstone}
\Phi_1=
\left(
\begin{array}{c}
-i(c_\beta w^+ - s_\beta H^+) \\
\frac{v_1 + \eta_1 + i(c_\beta z - s_\beta \eta_3)}{\sqrt{2}}
\end{array}
\right), \qquad
\Phi_2 =
\left(
\begin{array}{c}
-i(s_\beta w^+ + c_\beta H^+) \\
\frac{v_2 + \eta_2 + i(s_\beta z + c_\beta \eta_3 )}{\sqrt{2}}
\end{array}
\right),
\end{equation}
where $w^\pm$($z$) is the Goldstone boson of $W^\pm$($Z$). We remark that the $\eta_3$ pseudo-scalar field is orthogonal to the neutral Goldstone boson $z$.

As for the Goldstone boson mass spectrum, it is possible to find a
convenient way to write the mass matrix. Having each $\Phi_{1,2} $ the same group representation of the SM Higgs, following the notation of~\cite{Peskin:1995ev}, in the gauge-Goldstone\footnote{The
$4\times 3$ matrix follows from the four gauge bosons $W^i|_{i=1,3}$,
$Z$, and the three Goldstone bosons $\phi^i|_{i=1,3}$.} basis we find the following representation of the co-variant derivative:
\begin{equation}\label{gF}
\mathcal{D}_{1,2}=
\frac{v_{1,2}}{2}
\left(
\begin{array}{ccc}
g & 0 & 0 \\
0 & g & 0 \\
0 & 0 & g \\
0 & 0 & -g_1
\end{array} \right)\, .
\end{equation}

While the vector boson (and ghost) mass matrix is $m^2_V = \mathcal{D}_1 (\mathcal{D}_1)^T + \mathcal{D}_2 (\mathcal{D}_2)^T$, the Goldstones mass matrix is:
\begin{equation}
m^2_v = (\mathcal{D}_1)^T \mathcal{D}_1+(\mathcal{D}_2)^T \mathcal{D}_2\, ,
\end{equation}
therefore we get
\begin{equation}\label{GB_mass_matrix}
m^2_v = 
\frac{v^2}{4}
\left(
\begin{array}{ccc}
g^2 & 0	& 0 \\
0 & g^2	& 0 \\
0 & 0 & g^2+g_1^2 \\
\end{array} \right)\, .
\end{equation}

The mass matrix in Equation~(\ref{GB_mass_matrix}) shows that the
Goldstones have a mass that is equivalent to the SM-ones,
as expected since the gauge sector has not been extended.

As we have already intimated, the ghost mass matrix and interactions
are defined by means of the same matrix $\mathcal{D}$ via
\begin{eqnarray}
m^2_{ghost}=\mathcal{D}_1(\mathcal{D}_1)^T+\mathcal{D}_2(\mathcal{D}_2)^T.
\end{eqnarray}

Notice that the
$m^2_{ghost}$ and the $m^2_v$ of equation~(\ref{GB_mass_matrix}) have
different numbers of zero-eigenvalues, but their non-zero eigenvalues
are in a one-to-one correspondence; furthermore, the eigenvalues of
the gauge-fixing mass matrix are the same of the gauge boson mass
matrix, as expected.

Then, the ghost Lagrangian is defined as 
\begin{eqnarray}\label{L_gh}
\mathcal{L}_{ghost} &=& -\bar{c}^a \left[ (\partial_\mu D^\mu) ^{ab} +
\mathcal{D}_1^a \cdot \left( \mathcal{D}_1^b +\mathcal{S}_1^b \right)^T \right]
c^b\, \nonumber \\
 &-& \bar{c}^a \left[ \mathcal{D}_2^a \cdot \left( \mathcal{D}_2^b + \mathcal{S}_2^b \right)^T \right]
c^b\, ,
\end{eqnarray}
where the matrices $\mathcal{S}_{1,2}$ represent the link between the
fluctuations (Goldstones) of the Higgses around their VEVs and the
gauge bosons; a convenient\footnote{In fact, the explicit calculation of each single matrix gives rise to connection between ghosts and both the charged Higgses and the $\eta_3$ combination of scalar fields. Consistently, the sum of the two contributions disarms these connections: ghost fields do not link to charged Higgs.} way to write this matrices is
\begin{equation}
(\mathcal{S}_1)^T =
\frac{1}{2}\left(
\begin{array}{cccc}
g \eta_1 & c_\beta g z & -c_\beta g w_2 & -c_\beta g_1 w_2 \\
-c_\beta g z & g \eta_1 & c_\beta g w_1 & c_\beta g_1 w_1 \\
c_\beta g w_2 & -c_\beta g w_1 & g \eta_1 & -g_1 \eta_1 \\
\end{array}\right)\, ,
\end{equation}
\begin{equation}
(\mathcal{S}_2)^T =
\frac{1}{2}\left(
\begin{array}{cccc}
g \eta_2 & s_\beta g z & -s_\beta g w_2 & -s_\beta g_1 w_2 \\
-s_\beta g z & g \eta_2 & s_\beta g w_1 & s_\beta g_1 w_1 \\
s_\beta g w_2 & -s_\beta g w_1 & g \eta_2 & -g_1 \eta_2 \\
\end{array}\right)\, ,
\end{equation}
where
\begin{eqnarray}
w_1&=&\frac{w^++w^-}{\sqrt{2}}, \\
w_2&=&i\frac{w^+-w^-}{\sqrt{2}}.
\end{eqnarray}

Finally, the ghost fields ($\parbar{c}$) read as
\begin{equation}
c = \left( \begin{array}{cccc} 
w_1^g & w_2^g & w_3^g & B^g  
\end{array}\right)\, ,
\end{equation}
where
\begin{eqnarray}
w_1^g&=&\frac{w^+_g+w^-_g}{\sqrt{2}}, \\
w_2^g&=&i\frac{w^+_g-w^-_g}{\sqrt{2}}, \\
w_3^g&=& \cos{\theta_W}z_g+\sin{\theta_W}A_g, \\
B^g&=& -\sin{\theta_W}z_g+\cos{\theta_W}A_g.
\end{eqnarray}

\section{Perturbativity at high \boldmath$\tb$} 
\label{appe:perturbativity}

In order to properly realise the high $\tan\beta$ scenario, we must carefully consider the perturbativity of the couplings. Firstly, from Equation~(\ref{eq:M3}) we see that the $\tan\beta\to \infty$ approximation leads to the following definition of the heaviest neutral scalar mass:
\begin{eqnarray}
M_3^2=-\sum_{i=1}^2 M_i^2 \frac{R_{i2} R_{i3}}{R_{32} R_{33}}.
\end{eqnarray}

From Equations~(\ref{eq:lambda}) we can conclude that all the couplings are finite and perturbatively small except for $\lambda_1$ and $\lambda_3$.\footnote{While also $\im{\lambda_5}$ is apparently divergent, by plugging the definition of $M_3^2(\tan\beta\to \infty)$ in~(\ref{eq:lambda5I}) it is easy to see that the result is always finite.
If $v_1$ exactly vanishes, one can rotate away the phase of $\lambda_5$
to render $CP$-conservation manifest \cite{Ginzburg:2004vp}.
}
In fact, for $\tan\beta \to \infty$ or equivalently $\cos{\beta}\to 0$ and $\sin{\beta}\to 1$,
the two potentially diverging couplings become:
\begin{align}
\lambda_1&\simeq\frac{1}{v^2\cos^2{\beta}}\left[-\mu^2+\sum M_i^2 \left(R_{i1}^2-\frac{R_{i2} R_{i3} R_{31}^2}{R_{32} R_{33}}\right)\right],
\\
\lambda_3&\simeq\frac{M_1^2-M_2^2}{v^2\cos{\beta}}\,
\frac{R_{22} R_{12}}{R_{33}}.
\end{align}
Hence, no matter how large $\tb$ is,
we can always keep $\lambda_1$ and $\lambda_3$ perturbatively small
by choosing
\begin{gather}
\mu^2 \simeq \sum M_i^2 \left(R_{i1}^2-\frac{R_{i2} R_{i3} R_{31}^2}{R_{32} R_{33}}\right) ,
\\
M_1 \simeq M_2,
\end{gather}
respectively.  
This shows that
there is no strict upper limit on $\tb$ from perturbativity.
However, we do not make use of this mechanism
for drawing our conclusions in the main part of the present paper.


\section{\boldmath$p_T$ distributions of
$b$ and $\tau$}
\label{appe:diff}

\begin{figure}
  \centering
  \includegraphics[width=0.48\textwidth]{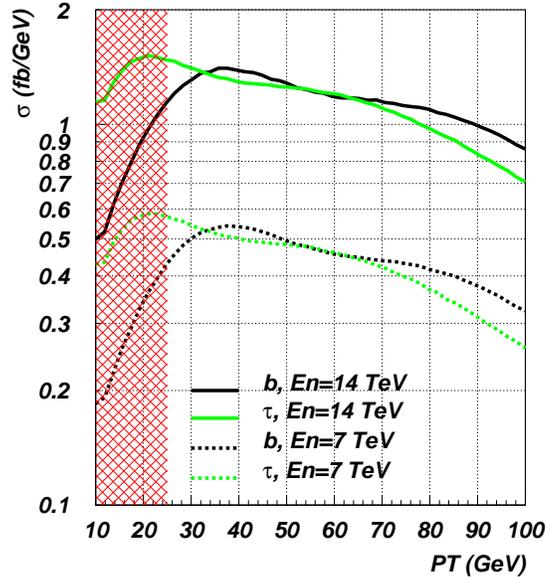}
\caption{Differential cross-sections of the process
$pp\rightarrow W \rightarrow H^\pm H_1 \rightarrow \tau^\pm \nunubarl{\tau} b\bar{b}$ with respect to $p_T$ 
of the $b$-quark and the $\tau$-lepton at the indicated
centre-of-mass energies, for the benchmark point $\Pa$.
Each curve is the sum of
$d\sigma(pp \rightarrow \tau^+ \nu b\bar{b})/d p_T$
and that for the conjugate final state.
The shadowed range (red online) shows the selection cuts
set for the signal-to-background analysis.}
\label{fig:spectra}
\end{figure}
In figure~\ref{fig:spectra},
we show the $p_T$ spectra for the $b$-quark and
the $\tau$-lepton, extracted from the process
$pp\rightarrow W \rightarrow H^\pm H_1 \rightarrow \tau^\pm \nu b\bar{b}$
at the LHC\@.
They were evaluated at the benchmark point $\Pa$.
In the $p_{T,b}$ distributions,
one finds that the selection cuts in~(\ref{eq:cuts}) act on regions
with suppressed differential cross-sections.
This means that we retain the major portion of events at our disposal
even after applying the cuts.
Even in the $p_{T,\tau}$ distributions, the excluded area below each curve
is relatively small compared to the total cross-section.






\begin{thebibliography}{10}

\bibitem{higgs mass naturalness}
S.~Weinberg, {\it {Gauge hierarchies}},  {\em Phys.\ Lett.}\  {\bf B82} (1979) 387;
M.~Veltman, {\it {The infrared--ultraviolet connection}},  {\em Acta
  Phys.\ Polon.}\  {\bf B12} (1981) 437; 
C.~H.~Llewellyn Smith and G.~G. Ross, {\it {The real gauge hierarchy problem}},
  {\em Phys.\ Lett.}\  {\bf B105} (1981) 38.

\bibitem{Fayet:1976et}
P.~Fayet, {\it {Supersymmetry and weak, electromagnetic and strong
  interactions}},  {\em Phys.\ Lett.}\  {\bf B64} (1976) 159.

\bibitem{Lee CPV}
T.~D.~Lee, {\it {A theory of spontaneous $T$ violation}},  {\em Phys.\ Rev.}\  {\bf D8}
  (1973) 1226;
{\it {$CP$ nonconservation and spontaneous symmetry breaking}},  {\em
  Phys.\ Rept.}\  {\bf 9} (1974) 143.

\bibitem{Weinberg:1976hu}
S.~Weinberg, {\it {Gauge theory of $CP$ violation}},  {\em Phys.\ Rev.\ Lett.}\  {\bf
  37} (1976) 657.

\bibitem{E6}
  See e.g.\
  J.~L.~Rosner, {\it
    An $E_6$ interpretation of an $e^{+} e^{-} \gamma \gamma\,\slashed{E}_T$ event},
  {\em Phys.\ Rev.}\  {\bf D55} (1997) 3143
  [\href{http://arxiv.org/abs/hep-ph/9607467}{hep-ph/9607467}];
S.~King, S.~Moretti, and R.~Nevzorov, {\it {Exceptional supersymmetric standard
  model}},  {\em Phys.\ Lett.}\  {\bf B634} (2006) 278,
  [\href{http://arxiv.org/abs/hep-ph/0511256}{{hep-ph/0511256}}].

\bibitem{Basso:2011na}
  See e.g.\
L.~Basso, S.~Moretti, and G.~M. Pruna, {\it {Theoretical constraints on the
  couplings of non-exotic minimal $Z'$ bosons}},  {\em JHEP} {\bf 1108} (2011)
  122, [\href{http://arxiv.org/abs/1106.4762}{{arXiv:1106.4762}}].

\bibitem{Peccei:1977ur}
R.~Peccei and H.~R. Quinn, {\it {Constraints imposed by $CP$ conservation in the
  presence of instantons}},  {\em Phys.\ Rev.}\  {\bf D16} (1977) 1791.

\bibitem{Antoniadis:2000ena}
  See e.g.\
I.~Antoniadis, E.~Kiritsis, and T.~Tomaras, {\it {A D-brane alternative to
  unification}},  {\em Phys.\ Lett.}\  {\bf B486} (2000) 186,
  [\href{http://arxiv.org/abs/hep-ph/0004214}{{hep-ph/0004214}}].

\bibitem{Gonderinger:2009jp}
  See e.g.\
M.~Gonderinger, Y.~Li, H.~Patel, and M.~J. Ramsey-Musolf, {\it {Vacuum
  stability, perturbativity, and scalar singlet dark matter}},  {\em JHEP} {\bf
  01} (2010) 053, [\href{http://arxiv.org/abs/0910.3167}{{arXiv:0910.3167}}].

\bibitem{Haber:1978jt}
H.~Haber, G.~L. Kane, and T.~Sterling, {\it {The fermion mass scale and
  possible effects of Higgs bosons on experimental observables}},  {\em
  Nucl.\ Phys.}\  {\bf B161} (1979) 493.

\bibitem{Babu:2004tn}
  See e.g.\
K.~S. Babu and J.~Kubo, {\it {Dihedral families of quarks, leptons and
  Higgses}},  {\em Phys.\ Rev.}\  {\bf D71} (2005) 056006,
  [\href{http://arxiv.org/abs/hep-ph/0411226}{{hep-ph/0411226}}].

\bibitem{neutrino mass}
  See e.g.\
E.~Ma, {\it {Naturally small seesaw neutrino mass with no new physics beyond
  the TeV scale}},  {\em Phys.\ Rev.\ Lett.}\  {\bf 86} (2001) 2502--2504,
  [\href{http://arxiv.org/abs/hep-ph/0011121}{{hep-ph/0011121}}];
S.~M. Davidson and H.~E. Logan, {\it {Dirac neutrinos from a second Higgs
  doublet}},  {\em Phys.\ Rev.}\  {\bf D80} (2009) 095008,
  [\href{http://arxiv.org/abs/0906.3335}{{arXiv:0906.3335}}].

\bibitem{Deshpande:1977rw}
N.~G. Deshpande and E.~Ma, {\it {Pattern of symmetry breaking with two Higgs
  doublets}},  {\em Phys.\ Rev.}\  {\bf D18} (1978) 2574.

\bibitem{idm}
R.~Barbieri, L.~J. Hall, and V.~S. Rychkov, {\it {Improved naturalness with a
  heavy Higgs: An alternative road to LHC physics}},  {\em Phys.\ Rev.}\  {\bf
  D74} (2006) 015007, [\href{http://arxiv.org/abs/hep-ph/0603188}{{hep-ph/0603188}}];
M.~Gustafsson, E.~Lundstrom, L.~Bergstrom, and J.~Edsjo, {\it {Significant
  gamma lines from inert Higgs dark matter}},  {\em Phys.\ Rev.\ Lett.}\  {\bf 99}
  (2007) 041301, [\href{http://arxiv.org/abs/astro-ph/0703512}{{astro-ph/0703512}}].

\bibitem{Goh:2009wg}
  See e.g.\
H.-S. Goh, L.~J. Hall, and P.~Kumar, {\it {The leptonic Higgs as a messenger of
  dark matter}},  {\em JHEP} {\bf 0905} (2009) 097,
  [\href{http://arxiv.org/abs/0902.0814}{{arXiv:0902.0814}}].

\bibitem{Turok:1990zg}
  See e.g.\
N.~Turok and J.~Zadrozny, {\it {Electroweak baryogenesis in the two doublet
  model}},  {\em Nucl.\ Phys.}\  {\bf B358} (1991) 471.

\bibitem{Gong:2012ri}
  See e.g.\
J.-O. Gong, H.~M. Lee, and S.~K. Kang, {\it {Inflation and dark matter in two
  Higgs doublet models}},  \href{http://arxiv.org/abs/1202.0288}{{arXiv:1202.0288}}. 

\bibitem{Schabinger:2005ei}
  See e.g.\
R.~Schabinger and J.~D. Wells, {\it {A minimal spontaneously broken hidden
  sector and its impact on Higgs boson physics at the Large Hadron Collider}},
  {\em Phys.\ Rev.}\  {\bf D72} (2005) 093007,
  [\href{http://arxiv.org/abs/hep-ph/0509209}{{hep-ph/0509209}}].

\bibitem{Cheung:2001hz}
K.-m.~Cheung, C.-H. Chou, and O.~C. Kong, {\it {Muon anomalous magnetic moment,
  two Higgs doublet model, and supersymmetry}},  {\em Phys.\ Rev.}\  {\bf D64}
  (2001) 111301, [\href{http://arxiv.org/abs/hep-ph/0103183}{{hep-ph/0103183}}].

\bibitem{Buras:2010mh}
A.~J. Buras, M.~V. Carlucci, S.~Gori, and G.~Isidori, {\it {Higgs-mediated
  FCNCs: natural flavour conservation vs.\ minimal flavour violation}},  {\em
  JHEP} {\bf 1010} (2010) 009, [\href{http://arxiv.org/abs/1005.5310}{{arXiv:1005.5310}}].

\bibitem{Park:2006gk}
J.-h.~Park, {\it {Lepton non-universality at LEP and charged Higgs}},  {\em
  JHEP} {\bf 0610} (2006) 077, [\href{http://arxiv.org/abs/hep-ph/0607280}{{hep-ph/0607280}}].

\bibitem{Ko:2011di}
P.~Ko, Y.~Omura, and C.~Yu, {\it {Chiral $U(1)$ flavor models and flavored Higgs
  doublets: the top FB asymmetry and the $Wjj$}},  {\em JHEP} {\bf 1201} (2012)
  147, [\href{http://arxiv.org/abs/1108.4005}{{arXiv:1108.4005}}]. 

\bibitem{ATLAS:2012ad}
{ATLAS Collaboration}, G.~Aad {\em et~al.}, {\it {Search for
  the Standard Model Higgs boson in the diphoton decay channel with $4.9$ fb$^{-1}$ of
  $pp$ collisions at $\sqrt{s}=7$ TeV with ATLAS}},  {\em Phys.\ Rev.\ Lett.}\  {\bf 108}
  (2012) 111803, [\href{http://arxiv.org/abs/1202.1414}{{arXiv:1202.1414}}].

\bibitem{Chatrchyan:2012tw}
{CMS Collaboration}, S.~Chatrchyan {\em et~al.}, {\it {Search
  for the Standard Model Higgs boson decaying into two photons in $pp$ collisions
  at $\sqrt{s}=7$ TeV}},  \href{http://arxiv.org/abs/1202.1487}{{arXiv:1202.1487}}.

\bibitem{ATLAS:2012ae}
{ATLAS Collaboration}, G.~Aad {\em et~al.}, {\it {Combined
  search for the Standard Model Higgs boson using up to 4.9 fb$^{-1}$ of $pp$
  collision data at $\sqrt{s} = 7$ TeV with the ATLAS detector at the LHC}},  {\em
  Phys.\ Lett.}\  {\bf B710} (2012) 49,
  [\href{http://arxiv.org/abs/1202.1408}{{arXiv:1202.1408}}].

\bibitem{Chatrchyan:2012tx}
{CMS Collaboration}, S.~Chatrchyan {\em et~al.}, {\it
  {Combined results of searches for the Standard Model Higgs boson in $pp$
  collisions at $\sqrt{s} = 7$ TeV}},  \href{http://arxiv.org/abs/1202.1488}{{arXiv:1202.1488}}.

\bibitem{Alcaraz:2006mx}
{ALEPH Collaboration, DELPHI Collaboration, L3 Collaboration, OPAL
  Collaboration, LEP Electroweak Working Group}, J.~Alcaraz {\em
  et~al.}, {\it {A combination of preliminary electroweak measurements and
  constraints on the Standard Model}},
  \href{http://arxiv.org/abs/hep-ex/0612034}{{hep-ex/0612034}}.

\bibitem{Dermisek light charged higgs}
R.~Dermisek, {\it {Light charged Higgs and lepton universality in $W$ boson
  decays}},  \href{http://arxiv.org/abs/0807.2135}{{arXiv:0807.2135}};
R.~Dermisek, {\it {Light charged and $CP$-odd Higgses in MSSM-like models}},
  {\em AIP Conf.\ Proc.}\  {\bf 1078} (2009) 226,
  [\href{http://arxiv.org/abs/0809.3545}{{arXiv:0809.3545}}];
K.~J. Bae, R.~Dermisek, D.~Kim, H.~D. Kim, and J.-H. Kim, {\it {Light Higgs
  scenario in BMSSM and LEP precision data}},
  \href{http://arxiv.org/abs/1001.0623}{{arXiv:1001.0623}}.

\bibitem{Filipuzzi:2012mg}
A.~Filipuzzi, J.~Portoles, and M.~Gonzalez-Alonso, {\it {$U(2)^5$ flavor symmetry
  and lepton universality violation in $W \to \tau \nu_\tau$}},
  \href{http://arxiv.org/abs/1203.2092}{{arXiv:1203.2092}}. 

\bibitem{Barger:1989fj}
  See e.g.\
V.~D. Barger, J.~Hewett, and R.~Phillips, {\it {New constraints on the charged
  Higgs sector in two Higgs doublet models}},  {\em Phys.\ Rev.}\  {\bf D41} (1990)
  3421.

\bibitem{Pilaftsis:1999qt}
  A.~Pilaftsis and C.~E.~M.~Wagner,
  {\it Higgs bosons in the minimal supersymmetric standard model with explicit $CP$ violation},
  {\em Nucl.\ Phys.}\  {\bf B553} (1999) 3,
  [\href{http://arxiv.org/abs/hep-ph/9902371}{{hep-ph/9902371}}].

\bibitem{Glashow:1976nt}
S.~L. Glashow and S.~Weinberg, {\it {Natural conservation laws for neutral
  currents}},  {\em Phys.\ Rev.}\  {\bf D15} (1977) 1958.

\bibitem{ElKaffas:2006nt}
A.~W. El~Kaffas, W.~Khater, O.~M. Ogreid, and P.~Osland, {\it {Consistency of
  the two Higgs doublet model and $CP$ violation in top production at the LHC}},
  {\em Nucl.\ Phys.}\  {\bf B775} (2007) 45,
  [\href{http://arxiv.org/abs/hep-ph/0605142}{{hep-ph/0605142}}].

\bibitem{Khater:2003wq}
W.~Khater and P.~Osland, {\it {$CP$ violation in top quark production at the LHC
  and two Higgs doublet models}},  {\em Nucl.\ Phys.}\  {\bf B661} (2003) 209,
  [\href{http://arxiv.org/abs/hep-ph/0302004}{{hep-ph/0302004}}].

\bibitem{Ginzburg:2005dt}
I.~Ginzburg and I.~Ivanov, {\it {Tree-level unitarity constraints in the most
  general 2HDM}},  {\em Phys.\ Rev.}\  {\bf D72} (2005) 115010,
  [\href{http://arxiv.org/abs/hep-ph/0508020}{{hep-ph/0508020}}].

\bibitem{Kanemura:2004mg}
See e.g.\
S.~Kanemura, Y.~Okada, E.~Senaha, and C.-P. Yuan, {\it {Higgs coupling
  constants as a probe of new physics}},  {\em Phys.\ Rev.}\  {\bf D70} (2004)
  115002, [\href{http://arxiv.org/abs/hep-ph/0408364}{{hep-ph/0408364}}].

\bibitem{Abbiendi:2008aa}
{OPAL Collaboration}, G.~Abbiendi {\em et~al.}, {\it {Search
  for charged Higgs bosons in $e^+e^-$ collisions at
  $\sqrt{s} = 189$--$209$ GeV}},
{\em Eur.\ Phys.\ J.}\ {\bf C72} (2012) 2076,
[\href{http://arxiv.org/abs/0812.0267}{{arXiv:0812.0267}}].

\bibitem{SM Higgs exclusion}
  ATLAS Collaboration, {\it
    An update to the combined search for the Standard Model
    Higgs boson with the ATLAS detector at the LHC using up to
    4.9 fb$^{-1}$ of $pp$ collision data at $\sqrt{s} = 7$ TeV},
  \href{http://cdsweb.cern.ch/record/1430033}{ATLAS-CONF-2012-019};
  CMS Collaboration, M.~Pieri {\em et~al.}, {\it
Searches for the Standard Model Higgs Boson at CMS},
\href{http://arxiv.org/abs/1205.2907}{arXiv:1205.2907}.

\bibitem{Haber:1999zh}
  H.~E.~Haber and H.~E.~Logan,
{\it Radiative corrections to the $Z b \overline{b}$ vertex and constraints on  extended
  Higgs sectors},
  {\em Phys.\ Rev.}\  {\bf D62}, 015011 (2000)
  [\href{http://arxiv.org/abs/hep-ph/9909335}{hep-ph/9909335}].

\bibitem{Tmasssplit}
  D.~Toussaint,
{\it Renormalization effects from superheavy Higgs particles},
{\em  Phys.\ Rev.}\  {\bf D18}, 1626 (1978);
  S.~Bertolini,
{\it Quantum effects in a two Higgs doublet model of the electroweak
  interactions},
{\em  Nucl.\ Phys.}\  {\bf B272}, 77 (1986);
  W.~Hollik,
{\it Nonstandard Higgs bosons in $SU(2) \times U(1)$ radiative corrections},
{\em  Z.\ Phys.}\  {\bf C32}, 291 (1986);
{\it Radiative corrections with two Higgs doublets at LEP / SLC and HERA},
{\em  Z.\ Phys.}\  {\bf C37}, 569 (1988).

\bibitem{LHC tbHc}
{ATLAS Collaboration}, G.~Aad {\em et~al.}, {\it {Search for
  charged Higgs bosons decaying via $H^+ \to \tau \nu$ in top quark pair events
  using $pp$ collision data at $\sqrt{s} = 7$ TeV with the ATLAS detector}},
  \href{http://arxiv.org/abs/1204.2760}{{arXiv:1204.2760}}; 
  {CMS Collaboration}, S.~Chatrchyan {\it et al.}, {\it
Search for a light charged Higgs boson in top quark decays in $pp$ collisions at $\sqrt{s} = 7$ TeV},
  \href{http://arxiv.org/abs/1205.5736}{{arXiv:1205.5736}}.

\bibitem{DiazCruz:1992gg}
  J.~L.~Diaz-Cruz and O.~A.~Sampayo,
  {\it Contribution of gluon fusion to the production of charged Higgs at hadron colliders},
  {\em Phys.\ Rev.}\  {\bf D50} (1994) 6820.

\bibitem{Djouadi:1999rca}
A.~Djouadi, W.~Kilian, M.~Muhlleitner, and P.~Zerwas, {\it {Production of
  neutral Higgs boson pairs at LHC}},  {\em Eur.\ Phys.\ J.}\  {\bf C10} (1999)
  45, [\href{http://arxiv.org/abs/hep-ph/9904287}{{hep-ph/9904287}}].

\bibitem{Akeroyd fermiophobic higgs}
A.~G. Akeroyd and M.~A. Diaz, {\it {Searching for a light fermiophobic Higgs
  boson at the Tevatron}},  {\em Phys.\ Rev.}\  {\bf D67} (2003) 095007,
  [\href{http://arxiv.org/abs/hep-ph/0301203}{{hep-ph/0301203}}];
A.~Akeroyd, M.~A. Diaz, and F.~J. Pacheco, {\it {Double fermiophobic Higgs
  boson production at the CERN LHC and LC}},  {\em Phys.\ Rev.}\  {\bf D70} (2004)
  075002, [\href{http://arxiv.org/abs/hep-ph/0312231}{{hep-ph/0312231}}].

\bibitem{Ginzburg:2012hc}
  I.~F.~Ginzburg,
  {\it Light charged Higgs at LHC},
\href{http://arxiv.org/abs/1205.5890}{{arXiv:1205.5890}}.

\bibitem{Semenov:1996es}
A.~V. Semenov, {\it {LanHEP: A package for automatic generation of Feynman
  rules in gauge models}},  \href{http://arxiv.org/abs/hep-ph/9608488}{{hep-ph/9608488}}.

\bibitem{calchep}
A.~Pukhov, {\it {CalcHEP 2.3: MSSM, structure functions, event generation, batchs, and generation of matrix elements for other packages}},
  \href{http://arxiv.org/abs/hep-ph/0412191}{{hep-ph/0412191}};
\href{http://theory.sinp.msu.ru/~pukhov/calchep.html}
{http://theory.sinp.msu.ru/~pukhov/calchep.html}.

\bibitem{Nadolsky:2008zw}
  P.~M.~Nadolsky, H.-L.~Lai, Q.-H.~Cao, J.~Huston, J.~Pumplin, D.~Stump, W.-K.~Tung and C.-P.~Yuan,
{\it Implications of CTEQ global analysis for collider observables},
{\em  Phys.\ Rev.}\  {\bf D78} (2008) 013004
  [\href{http://arxiv.org/abs/0802.0007}{arXiv:0802.0007}].

\bibitem{Hahn:1998yk}
T.~Hahn and M.~Perez-Victoria, {\it {Automatized one loop calculations in
  four-dimensions and $D$-dimensions}},  {\em Comput.\ Phys.\ Commun.}\  {\bf 118}
  (1999) 153, [\href{http://arxiv.org/abs/hep-ph/9807565}{{hep-ph/9807565}}].

\bibitem{def significance}
S.~I. Bityukov and N.~V. Krasnikov, {\it {On the observability of a signal
  above background}},  {\em Nucl.\ Instrum.\ Meth.}\  {\bf A452} (2000) 518;
S.~Bityukov, S.~Erofeeva, N.~Krasnikov, and A.~Nikitenko, {\it {Program for
  evaluation of significance, confidence intervals and limits by direct
  calculation of probabilities}},  in {\em Statistical problems in particle
  physics, astrophysics and cosmology: Proceedings of PHYSTAT05, Oxford, UK,
  12--15 Sep 2005} (L.~Lyons and M.~K. \"{U}nel, eds.), pp.~106--107, Imperial
  College Press, 2005.

\bibitem{TEVNPH:2012ab}
{TEVNPH (Tevatron New Phenomina and Higgs Working Group), CDF
  Collaboration, D0 Collaboration}, {\it {Combined CDF and D0
  search for Standard Model Higgs boson production with up to 10.0 fb$^{-1}$ of
  data}},  \href{http://arxiv.org/abs/1203.3774}{{arXiv:1203.3774}}.

\bibitem{gamma-gamma fusion}
N.~Bernal, D.~Lopez-Val, and J.~Sola, {\it {Single Higgs-boson production
  through $\gamma\gamma$ scattering within the general 2HDM}},  {\em Phys.\ Lett.}\ 
  {\bf B677} (2009) 39, [\href{http://arxiv.org/abs/0903.4978}{{arXiv:0903.4978}}];
D.~Lopez-Val and J.~Sola, {\it {Single Higgs-boson production at a
  photon-photon collider: general 2HDM versus MSSM}},  {\em Phys.\ Lett.}\  {\bf
  B702} (2011) 246, [\href{http://arxiv.org/abs/1106.3226}{{arXiv:1106.3226}}].

\bibitem{Posch:2010hx}
P.~Posch, {\it {Enhancement of $h \to \gamma \gamma$ in the two Higgs doublet
  model type I}},  {\em Phys.\ Lett.}\  {\bf B696} (2011) 447,
  [\href{http://arxiv.org/abs/1001.1759}{{arXiv:1001.1759}}].

\bibitem{Arhrib:2012ia}
A.~Arhrib, R.~Benbrik, and N.~Gaur, {\it {$H\to \gamma \gamma$ in inert Higgs
  doublet model}},  \href{http://arxiv.org/abs/1201.2644}{{arXiv:1201.2644}}. 

\bibitem{Peskin:1995ev}
M.~E. Peskin and D.~V. Schroeder, {\em {An introduction to quantum field
  theory}}, Addison-Wesley, 1995.

\bibitem{Ginzburg:2004vp}
I.~F. Ginzburg and M.~Krawczyk, {\it {Symmetries of two Higgs doublet model and
  $CP$ violation}},  {\em Phys.\ Rev.}\  {\bf D72} (2005) 115013,
  [\href{http://arxiv.org/abs/hep-ph/0408011}{{hep-ph/0408011}}]. 

\end{thebibliography}

\end{document}

Local Variables:
mode: latex
TeX-source-correlate-method: source-specials
TeX-source-correlate-start-server: t
eval: (TeX-source-correlate-mode t)
End: